\documentclass[fleqn,usenatbib]{mnras}

\usepackage{newtxtext,newtxmath}

\usepackage[T1]{fontenc}

\DeclareRobustCommand{\VAN}[3]{#2}
\let\VANthebibliography\thebibliography
\def\thebibliography{\DeclareRobustCommand{\VAN}[3]{##3}\VANthebibliography}

\usepackage{graphicx}	
\usepackage{amsmath}	
\usepackage[normalem]{ulem}

\title[Variability and QPO evolution]{
Understanding corona and disk evolution in black hole X-ray binaries through a comprehensive study of their broadband variability and QPO characteristics 
}

\author[B. Ram et al.]{
Biki Ram,$^{1}$\thanks{E-mail: bikiram436@gmail.com}
Manoneeta Chakraborty,$^{1}$
\\
$^{1}$Department of Astronomy, Astrophysics and Space Engineering, Indian Institute of Technology Indore, Indore, 453552, India\\
}

\date{Accepted XXX Received YYY; in original form ZZZ}

\pubyear{\the\year{}}

\begin{document}
\label{firstpage}
\pagerange{\pageref{firstpage}--\pageref{lastpage}}
\maketitle

\begin{abstract}
The shape of the power spectrum of the black hole low-mass X-ray binary evolves systematically over different spectral states during an outburst. Therefore, the power colours (ratio of the variability amplitude at different frequency ranges) and the ‘hue’ parameter, quantifying the power spectral shape, can be utilised to identify the spectral states of the system. We present the comprehensive power colour analysis and subsequent identification of spectral states using the entire archival data (2016-2024) from \emph{AstroSat}. We detected 29 QPOs (quasi-periodic oscillations), along with several associated harmonics and shoulders, and investigated their properties as a function of hue. We examined the evolution of the QPO RMS variability and time lag, along with hue and QPO frequency. We report a sign change in the average QPO time lag around the QPO frequency of $\sim$2 Hz for high inclination sources, during the hard-to-hard intermediate state transition. At lower frequency, the hard lags showed an increasing trend reaching up to $\sim$36 ms, but the soft lags above 2 Hz remained confined within $\sim$10 ms, suggesting an evolution to a compact corona. Conversely, for low inclination sources, no such transition was found.  
Furthermore, for high inclination sources, the harmonic lag remains unaffected during state transition, in contrast to the QPO lag behaviour. 
Our results are consistent with a transition from an elongated jet-like corona to a compact corona and reveal vital clues about the dynamical evolution of the corona and disk.

\end{abstract}







\section{Introduction}
\label{introduction}

Black hole low-mass X-ray binaries (BH-LMXB) are remarkable astrophysical laboratories for studying the laws of physics under extreme gravity. 
These systems offer critical insights into accretion processes and associated high-energy phenomena through their emission spectra and variability patterns. 
During an outburst, the luminosity and observed properties of these sources change significantly as they traverse through different spectral states \citep{2006ARA&A..44...49R}.  
The soft photons from a geometrically thin and optically thick accretion disk in the black hole X-ray binaries (BH-XRBs) \citep{1973A&A....24..337S} interact with the comptonising component (corona) and gain energy through inverse comptonisation, producing the observed hard X-ray emission \citep{1976ApJ...204..187S,1980A&A....86..121S}.
However, the geometry of the corona and its variation over different spectral states still remain very ambiguous.   
The evolution of the BH-XRBs through the various spectral states is often visualised through a Hardness-Intensity Diagram (HID).
The HID for a BH-XRB typically displays a characteristic q-shaped track, as the source traverses from Low-Hard State (LHS) to Hard-Intermediate State (HIMS), then to Soft-Intermediate State (SIMS), and eventually reaches High-Soft State (HSS). Subsequently, again through SIMS and HIMS, it returns to the hard state at the end of the outburst \citep{2004MNRAS.355.1105F}.
The Hardness–Intensity Diagram (HID), though widely used to track spectral evolution during outbursts, has certain limitations. It is not suitable for comparing different sources within a single diagram, as their intrinsic properties, such as mass, distance, and inclination, affect both the emission level and spectral shape. 
Moreover, the HID is strongly instrument-dependent, and differences in detector response and effective area across instruments can lead to discrepancies in the observed tracks.

The timing properties of BH-XRBs change as the source goes through these different states. In the LHS, the energy spectrum is dominated by hard X-ray photons. 
The power spectrum exhibits strong broadband variability with flat-top noise across a wide range of frequencies with a fractional root mean square (RMS) variability amplitude of 30-50\% \citep{2005A&A...440..207B, 2006ARA&A..44...49R, 1997ApJ...485L..37M}. As the source goes towards the HIMS state, the spectrum gradually softens due to an enhanced contribution from the accretion disk. The system shows complex behaviour during the hard and the hard to HIMS transitional phase, marked by the emergence of quasi-periodic oscillations (QPOs)  \citep{1999ApJ...520..262P, 2000MNRAS.318..361N, 2005A&A...440..207B}. During this phase, the broadband noise becomes band-limited around a few Hz.
Further evolution leads to SIMS, where the broadband variability diminishes significantly, and type-C QPOs are often replaced by type-B QPOs. The fractional RMS becomes around 5-20\% \citep{2010LNP...794...53B, 2012MNRAS.427..595M} and finally, in the HSS state, it goes below 5\% \citep{2022hxga.book....9K,2024JApA...45...30R}. Thus, depending on the evolution of the power spectra and fractional amplitude, the spectral states of the BH-XRB  can be identified during an outburst.
A new method in the temporal domain is used to quantify the evolution of power spectral shapes across states, called the power-colour diagram. \citet{2015MNRAS.448.3339H} explored this by estimating the variance in four distinct frequency bands, using them to give a quantitative framework for distinguishing the different states of black hole LMXBs. \cite{2018MNRAS.481.3761G} used the same method to compare the variability of the neutron star and BH-LMXB sources. 
The power-colour method works as an alternative to the traditional HID method to identify different states.

Type-C QPO is frequently observed during the LHS and HIMS states \citep{1985Natur.313..768V,2019NewAR..8501524I} and is also occasionally detected in the soft and ultraluminous states (ULS) \citep{2012MNRAS.427..595M}.
The physical mechanism behind this feature is still debatable, but there are few promising mechanisms: relativistic precession model \citep{1998ApJ...492L..59S, 2006ApJ...642..420S,2012MNRAS.419.2369I},  instabilities of the corona–disk \citep{2004ApJ...612..988T, 2022A&A...662A.118M}, accretion–ejection instability mechanism  \citep{2022ApJ...930...18W}, disk instability model, blob-model, and variable corona model. \cite{10.1111} proposed Lense-Thirring to be the most favourable model for type-C QPOs involving a truncated disk geometry. However, none of these models is completely able to explain the observed characteristics of the QPOs comprehensively. 
Studying the behaviour of the QPO and its correlation with key spectral parameters enables a better understanding of the geometry of the QPO emitting region, offering valuable clues about both the physical origin of the QPOs and the evolution of the accretion geometry during state transitions.
For example, the inner disk radius shows an inconsistent behaviour with the QPO frequency, exhibiting different (increasing/decreasing) trends for different BH-XRB systems \citep{2000ApJ...531..537S}.  
However, the strong correlation between QPO frequency and the non-thermal component suggests that the corona might play a significant role in shaping QPO characteristics \citep{2003A&A...397..729V}. Recently, LFQPOs found at higher energies ( $>$ 100 keV) further strengthen the non-thermal connection of QPO \citep{2021NatAs...5...94M,2024MNRAS.531.1149N}.

The energy-dependent behaviour of the QPO properties can serve as a key diagnostic for the contributions and dynamical evolution of the accretion geometry and Comptonising region in BH-XRBs. 
The phase lag between hard and soft photons at the QPO frequency often manifests distinct transitions in its evolution, providing crucial insights into the evolving interaction between the disk and the corona in BH-XRBs. 
A transition in the phase or time-lag behaviour around a few hertz has been reported in multiple sources. \cite{2000ApJ...541..883R} first observed a decrease and sign reversal of phase lags $\sim$ 2 Hz in GRS 1915+105, later confirmed using all RXTE data by \citet{2020MNRAS.494.1375Z}. Similar breaks near 2 Hz and 3.3 Hz were also found in GX 339-4 and MAXI J1535-571, respectively \citep{2017ApJ...845..143Z, 2022MNRAS.512.2686Z}.
This specific nature of the transition of lag at a particular QPO frequency raises important questions about the physical origin of QPO, indicating complex mechanisms underlying their generation. 
Currently, the exact physical reason behind the systematic transition in QPO behaviour at this specific QPO frequency remains unclear.

\cite{2017MNRAS.464.2643V} later reported a strong inclination-dependence in the phase-lag behaviour of QPOs.
This result implies that the QPO phenomenon is strongly influenced by geometrical effects related to the inclination of the LMXB.
Inclination-dependent variations in emission properties have been reported across various accretion states \citep{2013ApJ...778..165V,2015MNRAS.448.3348H,2015MNRAS.447.2059M,2019A&A...625A..90R,2020MNRAS.491..313A}. 

QPOs are often accompanied by their harmonic components in the power spectrum.
Interestingly, the fundamental QPO and its harmonic exhibit distinct lag evolution. While the second harmonic always shows a hard lag, the lag associated with the fundamental QPO evolves systematically \citep{2005ApJ...629..403C,2017MNRAS.464.2643V,2013ApJ...778..136P,2013ApJ...779...71M,2020MNRAS.494.1375Z}. This raises another key scientific question: whether the fundamental QPO and its harmonic share a common physical origin or arise from distinct yet coupled mechanisms within the accretion flow.

In this study, we perform a comprehensive timing analysis of 14 black hole X-ray binaries (BH-XRBs) with extensive observational coverage from \emph{AstroSat}. 
While power-colour diagnostics have been successfully applied using data from \emph{Rossi X-ray Timing Explorer (RXTE)} and \emph{Neutron star Interior composition Explorer (NICER)} \citep{2015MNRAS.448.3339H,2022ApJ...930...18W}, no such systematic study has yet been carried out with \emph{AstroSat}, whose broad energy coverage and excellent timing resolution make it an ideal instrument to explore accretion variability across different spectral states.
We conducted a comprehensive analysis of the variability behaviour, the associated QPOs and harmonics, and their evolution with the spectral state variation.
We implemented a framework linking timing properties to spectral states, and traced the evolution of QPO characteristics as a function of hue. We performed the time-lag and RMS analysis for both high- and low-inclination sources to investigate the evolution of the accretion flow and corona. We further compare the timing properties of QPOs and their harmonics. Finally, we discuss our results, highlighting the role of hue in characterizing spectral state transitions 
and the evolution of timing behaviour with hue in BH-XRBs.

\begin{table}
\begin{center}
\scalebox{0.57}{%
    \begin{tabular}{ccccccc}
    \hline
    Source & Outburst & Obs ID & Start date &Start Time &  Exposure  \\
    & year&  & (dd-mm-yyyy)& (hh:mm:ss)& (ks) \\\hline
    H 1743 - 332&2016& 9000000364&09-03-2016&09:46:03&11.67 \\
    &&9000000562&27-07-2016&13:33:58&52.24 \\
    &2017&9000001444&08-08-2017&07:42:11&12.95\\
    \hline
    GRS 1716-249&2017&9000001034&15-02-2017&00:43:33&35.42 \\
    &&9000001140&06-04-2017&15:32:07&3.01 \\
    &&9000001378&13-07-2017&12:41:15&9.98\\
    \hline

    GRS 1758-258&2016&9000000732&15-10-2016&18:09:36&29.30 \\
    &2017&9000001410&28-07-2017&22:32:25&13.51 \\
    &&9000001542&21-09-2017&02:10:36& 36.17\\
    &2022&9000005250&29-07-2022&13:01:30&33.87 \\
    &&9000005358&11-10-2022&09:13:22& 27.78
    \\\hline
     Swift J1658.2-4242&2018&9000001910&21-02-2018&04:00:05&19.94\\
    &&9000001940&04-03-2018&06:21:45&29.59\\
     \hline
     MAXI J1803-298&2021&9000004368&11-05-2021&01:09:07&12.14
     \\
     &&9000004370&11-05-2021&12:05:30&45.94\\\hline
    GRS 1915+105 &2016&9000000358&04-03-2016&12:50:10& 68.02 \\
    &&9000000428&25-04-2016&05:15:11&35.64\\
    &&9000000432&27-04-2016&17:22:25&14.44\\
    &&9000000492&15-06-2016&03:55:04&26.94\\
    &&9000000546&20-07-2016&08:57:51&11.66\\
    &&9000000552&23-07-2016&09:33:36&23.18\\
    &&9000000574&03-08-2016&18:37:49&11.99\\
    &&9000000580&06-08-2016&09:08:07&10.58\\
    &&9000000652&10-09-2016&05:17:06&76.85\\
    &&9000000684&25-09-2016&16:10:22&75.54\\
    &&9000000760&28-10-2016&19:56:39&38.82\\
    &&9000000770&02-11-2016&11:16:20&12.55\\
    &&9000000792&13-11-2016&22:13:46&37.00\\
    &2017
    &9000001124&01-04-2017&11:50:10&10.08\\
    &&9000001162&14-04-2017&22:12:19&18.72\\
    &&9000001166&16-04-2017&00:08:08&11.69\\
    &&9000001232&18-05-2017&18:35:53&8.96\\
    &&9000001236&19-05-2017&21:40:52&20.42\\
    &&9000001272&05-06-2017&18:24:38&9.79\\
    &&9000001500&30-08-2017&02:15:08&10.16\\
    &&9000001506&31-08-2017&15:23:45&9.41\\
    &&9000001534&11-09-2017&13:38:56&18.68\\
    &&9000001618&15-10-2017&20:21:53&15.07\\
    &&9000001622&20-10-2017&08:40:17&11.75\\
    &&9000001630&21-10-2017&19:15:02&16.56\\
    &&9000001656&31-10-2017&18:25:43&15.37\\

    &2018&9000002000&01-04-2018&16:50:14&19.52\\
    &&9000002006&04-04-2018&02:28:21&16.58\\
    &&9000002080&08-05-2018&05:08:31&15.72\\
    &&9000002110&21-05-2018&16:25:16&84.09\\
    &&9000002112&24-05-2018&17:38:10&13.08 \\
    &&9000002190&26-06-2018&00:00:01&7.64\\
    &&9000002220&14-07-2018&17:05:50&38.17 \\
    &&9000002306&14-08-2018&10:31:23&16.74\\
    &&9000002334&28-08-2018&02:35:18&12.41\\
    &2019&9000002812&21-03-2019&16:34:31&79.84\\
    \hline
     
    GX 339-4 
    &2017&9000001578&05-10-2017&04:04:08&22.79 \\
    &2021&9000004180&13-02-2021&17:47:15&27.04 \\
    &&9000004218&04-03-2021&16:07:09&99.44\\     
    &&9000004222&05-03-2021&08:06:15&11.40\\
    &&9000004224&05-03-2021&09:10:32&1.63\\
    &&9000004226&05-03-2021&10:52:38&1.48\\
    &2022&9000005334&07-09-2022&08:06:45&29.15 \\
    &&9000005338&09-09-2022&08:37:19&89.67 
    \\
    &2024& 9000006070&15-02-2024&21:16:02&8.7\\\hline
    4U 1630-472 &2016&9000000626&27-08-2016&21:37:25&38.95 \\
    &&9000000686&29-09-2016&05:14:40&37.89 \\
    &&9000000698&02-10-2016&05:20:40&39.60 \\
    &2018&9000002274&04-08-2018&08:48:23&38.56  \\ 
    &&9000002282&07-08-2018&06:54:36&10.19  \\
    &&9000002286&09-08-2018&04:19:34&18.06 \\
    &&9000002294&10-08-2018&08:21:59&9.42 \\
    &&9000002298&11-08-2018&19:58:06&8.48 \\
    &&9000002304&14-08-2018&04:12:51&10.18 \\
    &&9000002318&23-08-2018&00:51:21&77.32 \\
    &&9000002340&30-08-2018&18:52:37&10.37\\
    &&9000002354&11-09-2018&10:54:57&10.32 \\
    &&9000002372&17-09-2018&03:51:07&9.88\\
    &2020
    &9000003612&15-04-2020&02:05:20&18.75 \\
    \hline
    MAXI J1348-630&2019&9000002722&19-02-2019&22:10:52&5.41 \\
    &&9000002728&22-02-2019&12:18:19&20.21 \\
    &&9000002742&28-02-2019&09:24:32&23.16\\ 
    &&9000002896&08-05-2019&19:24:02&13.47\\
    &&9000002990&14-06-2019&15:16:07&35.05
    \\\hline
    4U 1820+070 &2018&9000001994&31-03-2018&13:12:07&38.50 \\
    &&9000002216&10-07-2018&01:01:39&83.47 \\
    &&9000002324&25-08-2018&11:10:02&11.32  \\  
    \hline
    1E 1740-2942&2016&9000000714&07:10:2016&13:34:08&24.71\\
    &2018&9000001920&26-02-2018&11:07:39&29.31\\
    &&9000002092&11-05-2018&20:01:22&20.33\\
    &&9000002096&13-05-2018&18:54:20&35.64\\\hline
    Swift J151857-572147& 2024&9000006126&18-03-2024&17:03:27& 19.86\\\hline
    XTE  J1908+094&2018&9000002204&01-07-2018&08:11:34&109.32\\\hline
    IGR J17091-3624&2016&9000000430&26-04-2016&06:13:07&125.76\\
    &2022&9000005050&01-04-2022&04:09:46&92.76\\\hline
\end{tabular}}
\caption{Details of \emph{AstroSat} observations of all the outbursts of all the sources considered in this work. The observation ID is represented here. \label{Tab:1}}
\end{center}
\end{table}

\section{Observations and data reduction }\label{Sec:2}
We have used data from the \emph{AstroSat} satellite, which is the first Indian multi-wavelength satellite dedicated to astronomy. 
It has 5 instruments onboard: LAXPC (Large Area X-ray Proportional Counter), SXT (Soft X-ray Telescope), UVIT (Ultra-Violet Imaging Telescope), CZTI (Cadmium-Zinc-Telluride Imager), and SSM (Scanning Sky Monitor) \citep{2017JApA...38...30A}. 
For this study, we used the data from the LAXPC detector only. Our work involves a detailed temporal investigation of a sample of BH-XRB sources. Therefore, LAXPC, being one of the best timing instruments available, is the ideal instrument for conducting this study. It consists of three gas-filled (90 \% Xenon + 10 \% Methane) proportional counter units LAXPC10, LAXPC20, and LAXPC30 \citep{2017CSci..113..591Y,2017JApA...38...30A}. Shortly after launch, LAXPC30 experienced a leakage of gas; thus, it has been excluded from scientific analysis. After 2018, LAXPC10 also showed some signs of gas leakage and unstable gain. Therefore, for consistency, we have considered only the LAXPC20 data for all observations spanning a duration from 2016 to 2024. The timing resolution of LAXPC is 10 $\mu$s with the dead time of 42 $\mu$s. It covers an energy range of 3-80 keV, and has an effective area of 6000 cm$^2$/keV at 15 keV \citep{2017CSci..113..591Y,2017JApA...38...30A}. The energy resolution of the instrument is 15-20 \% at 30 keV. We have considered data till 30 keV energy range, as beyond that the background starts to dominate in most cases, reducing the reliability of the source signal. We considered the event mode data of LAXPC20 for this work.

Initially, we considered the data of all black hole X-ray binaries (BH-XRBs) observed by \emph{AstroSat}.
We used the LAXPC FORMAT-A\footnote[1]{http://astrosat-ssc.iucaa.in/laxpcData} software to create the level2 data from level 1 data obtained from the \emph{AstroSat} Data Archive\footnote[2]{https://astrobrowse.
issdc.gov.in/astroarchive/archive/Home.jsp}. We combined all the orbit file data for each observation using the {\tt laxpc\_make\_filelist} command. Then we created good time interval files (GTI files) using {\tt laxpc\_make\_stdgti} to exclude the effects of earth occultation and the South Atlantic Anomaly (SAA) from the level2 data. After the GTI selection, we created the light curve using {\tt laxpc\_make\_lightcurve}. 
For many light curves, even following GTI selection, there remain certain artefacts in the light curves that we removed by further modifying the GTI files to exclude the times of these features. Using these modified GTI files, we generated the spectra and background spectra using {\tt laxpc\_make\_spectra} and {\tt laxpc\_make\_backspectra} tools.
Subsequently, we used the {\tt laxpc\_find\_freqlag} tool to generate the power spectra considering a Nyquist frequency of 16 Hz and an integration time of 256 s. We excluded observations where the light curves were suboptimal, being filled with numerous data gaps post-GTI selection, and also those light curves with exposure of less than 1000 s. We also neglected observations and sources that have very noisy and weakly detectable variability in their power spectra. 
In cases where such behaviour was persistent throughout the observation, the entire dataset was discarded. After applying these selection criteria, we retained data from 14 BH-LMXB sources for which data were publicly available on the \emph{AstroSat} Data Archive. In table~\ref {Tab:1} we listed the sources selected in our sample along with the details of their corresponding \emph{AstroSat} observations. We have provided the related outburst years, observation ID, starting date, starting time of observation, and total exposure time in this table. 
 
\begin{figure*}
\includegraphics[scale=0.27]{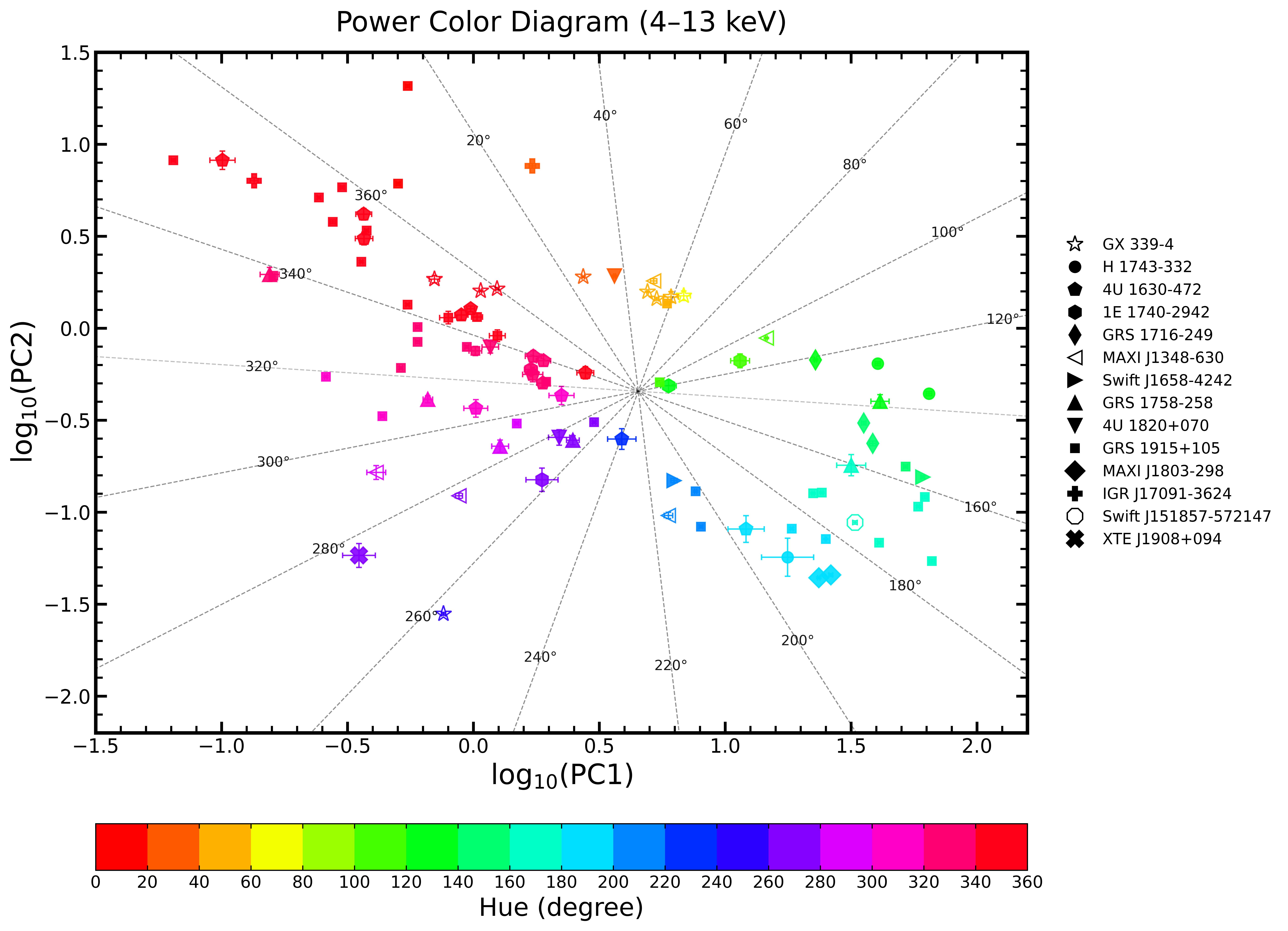}
\includegraphics[scale=0.27]{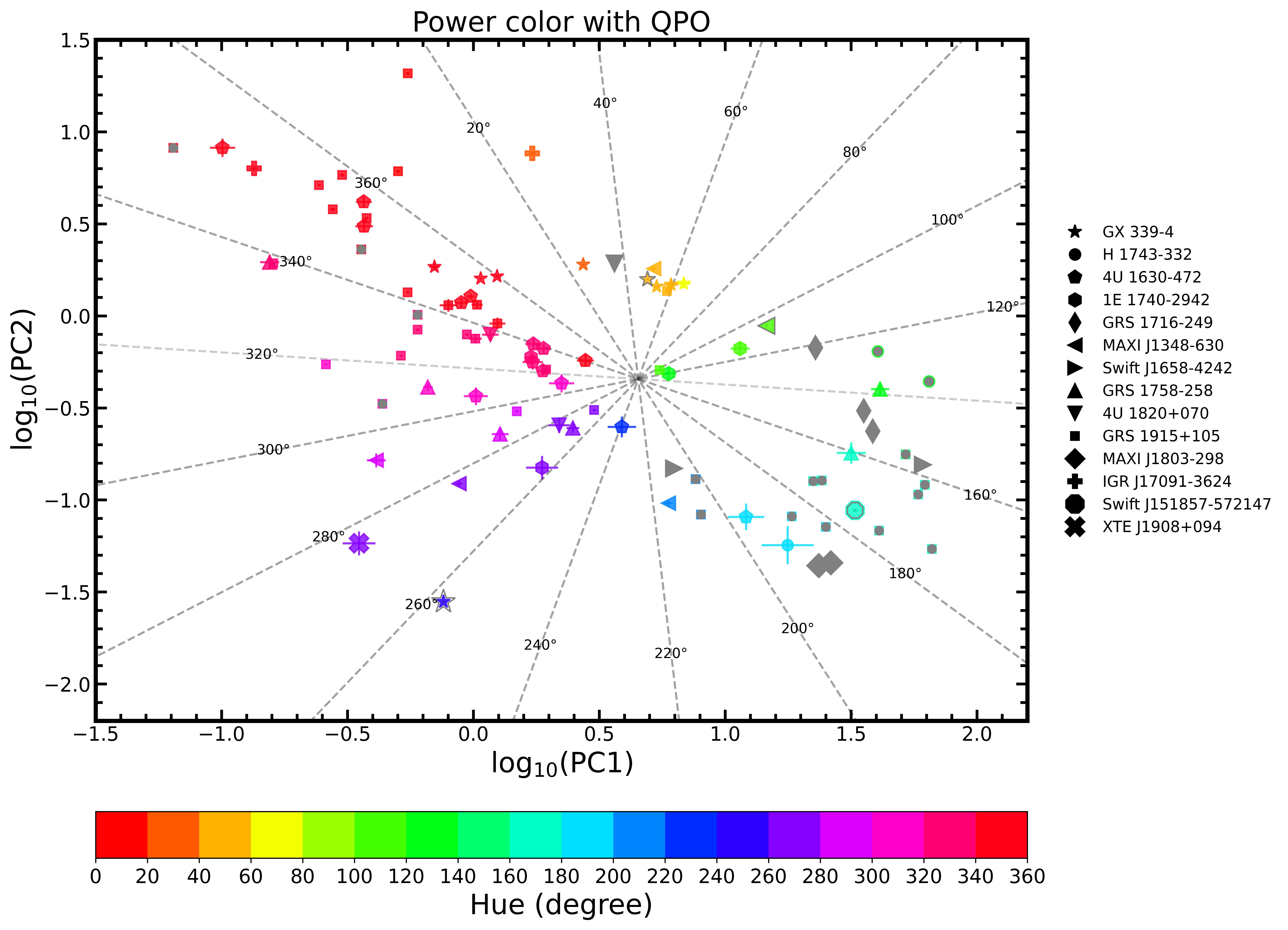}
\caption{Left: Power colour diagram for all the observations of all the sources considered in this work. Each source is represented by a distinct marker, as indicated in the legend, and the open markers represent the low inclination sources. The colour bar represents the power colour hue value (see Section \ref{Sec:3.1}) associated with the different observations. Hue intervals of $20^\circ$ are marked in the plot with grey dashed lines. Right plot: The same plot as the left one, with the observations comprising QPOs marked by the grey data points. The solid grey markers represent the high inclination sources with QPO detections, and the open grey markers represent the low inclination sources with QPO detections. \label{Fig:1} }
\end{figure*}

\begin{figure*}
\includegraphics[scale=0.40]
{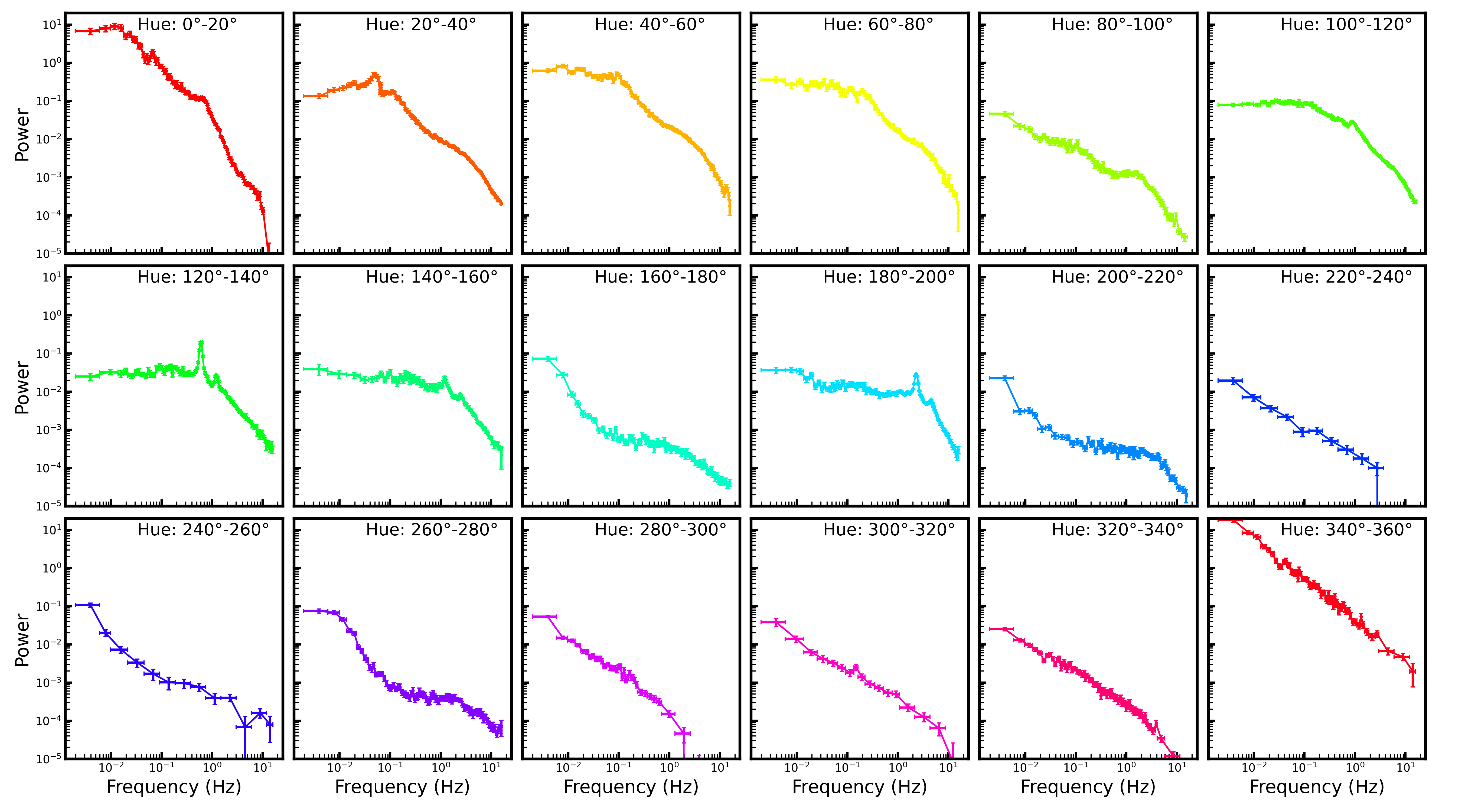}
\caption{ Evolution of the RMS normalised Poisson power subtracted power spectra in different hue bins. We have chosen one representative power spectrum in each hue bin of 20. Different colours represent different hue regions on the power colour wheel as depicted in Figure~\ref{Fig:1}. The hue values corresponding to each representative power spectra plot are mentioned in each respective panel. \label{Fig:2}}
\end{figure*}

\begin{figure*}
\begin{center}
\includegraphics[scale=0.08]{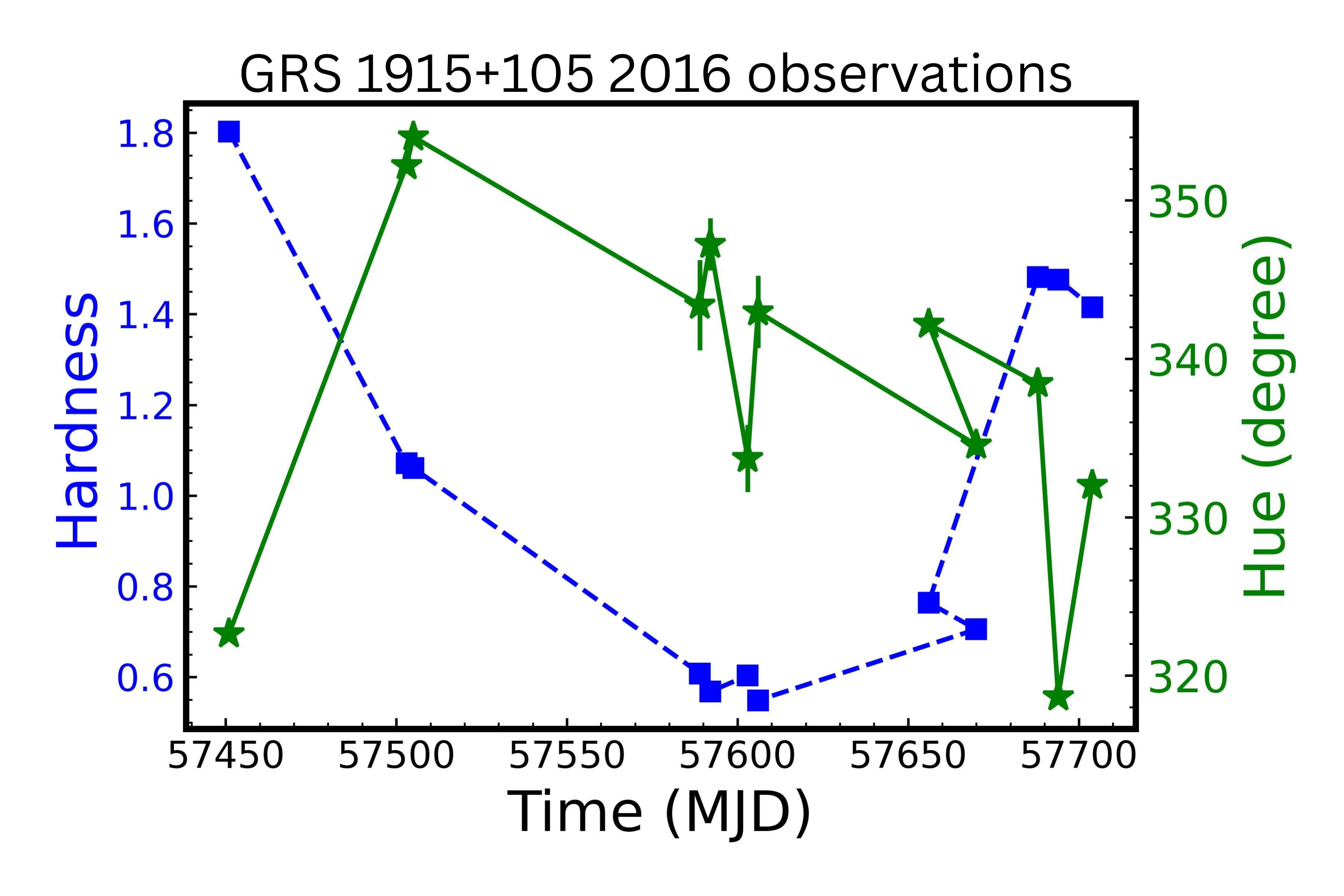}
\includegraphics[scale=0.08]{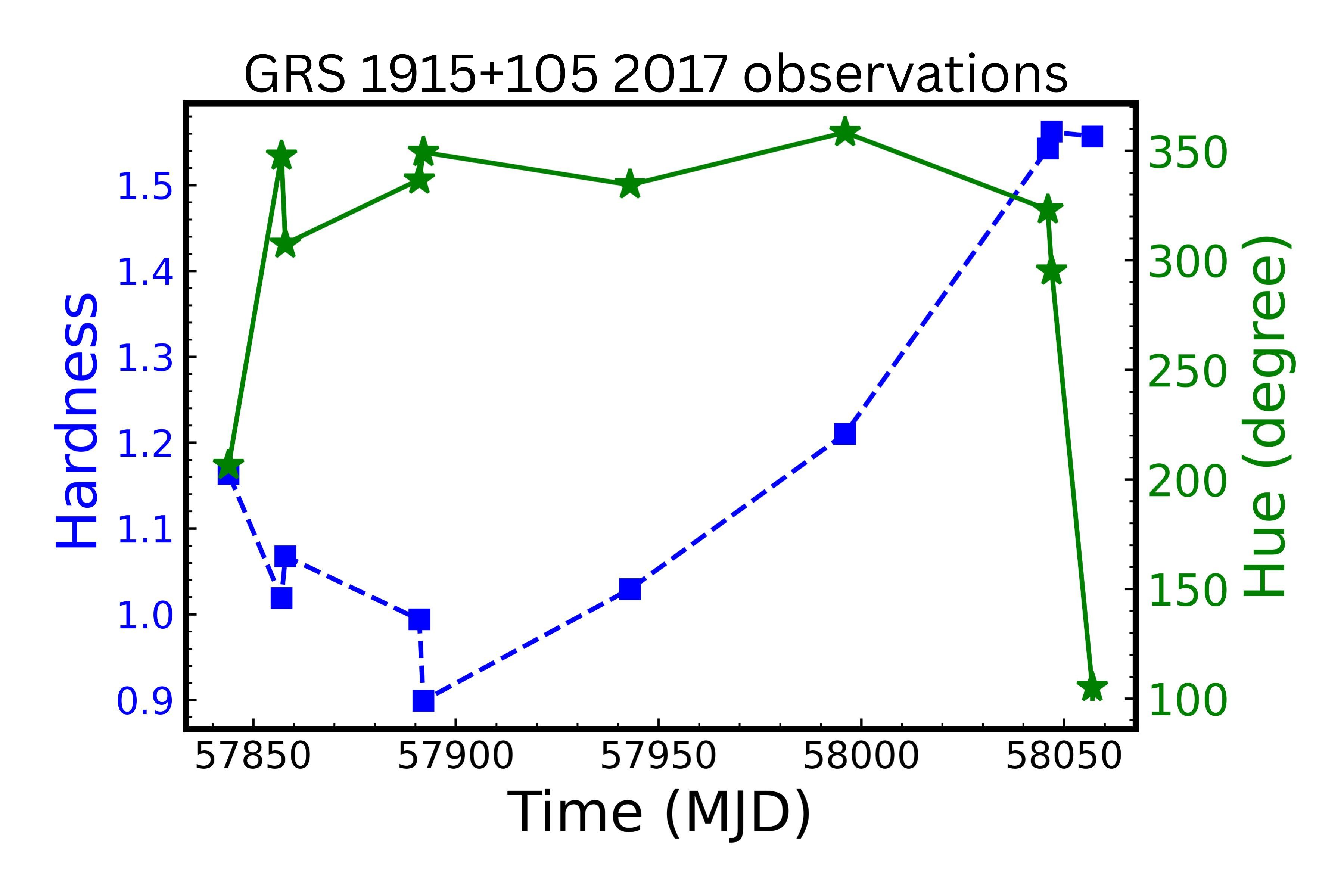}
\includegraphics[scale=0.08]{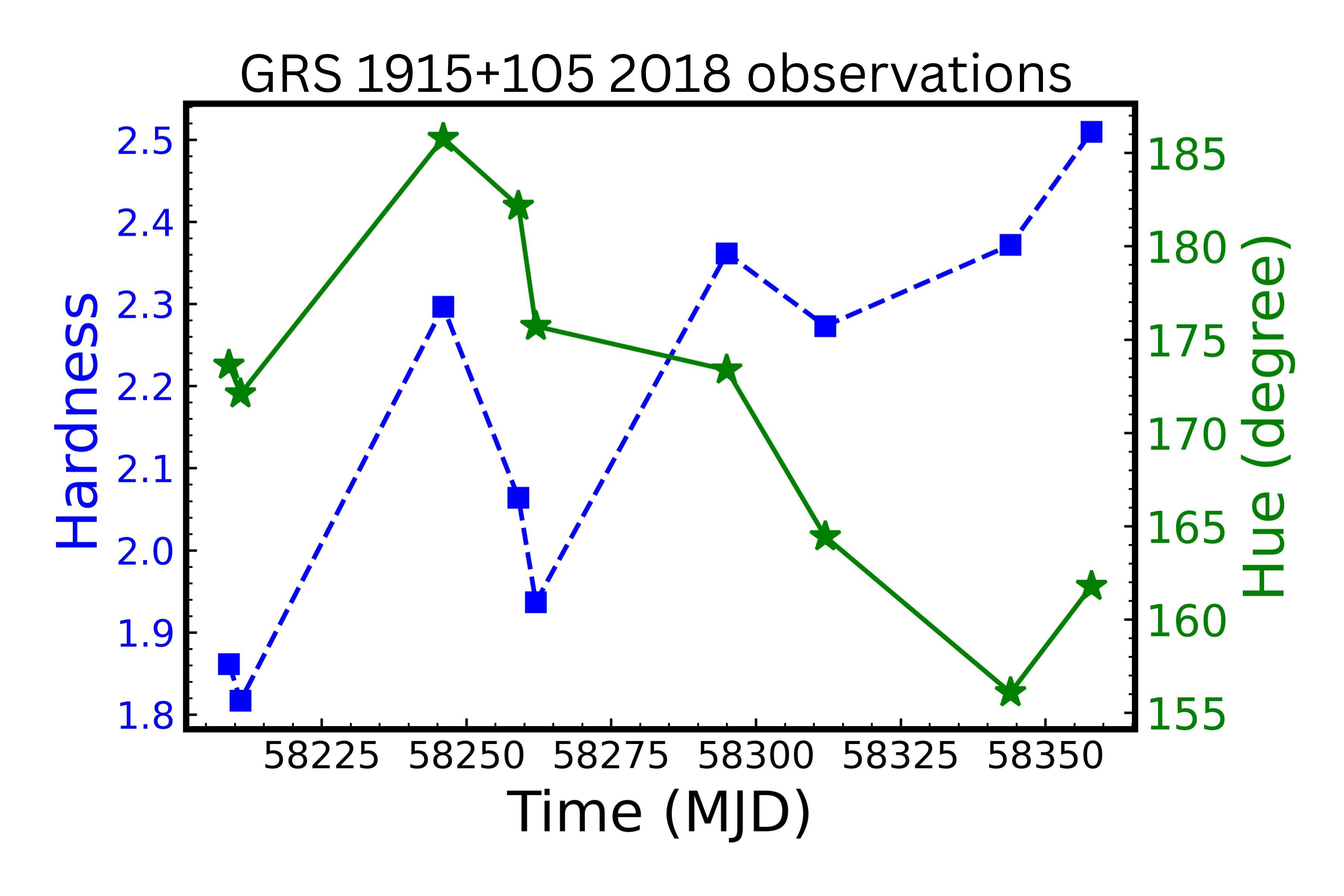}
\includegraphics[scale=0.08]{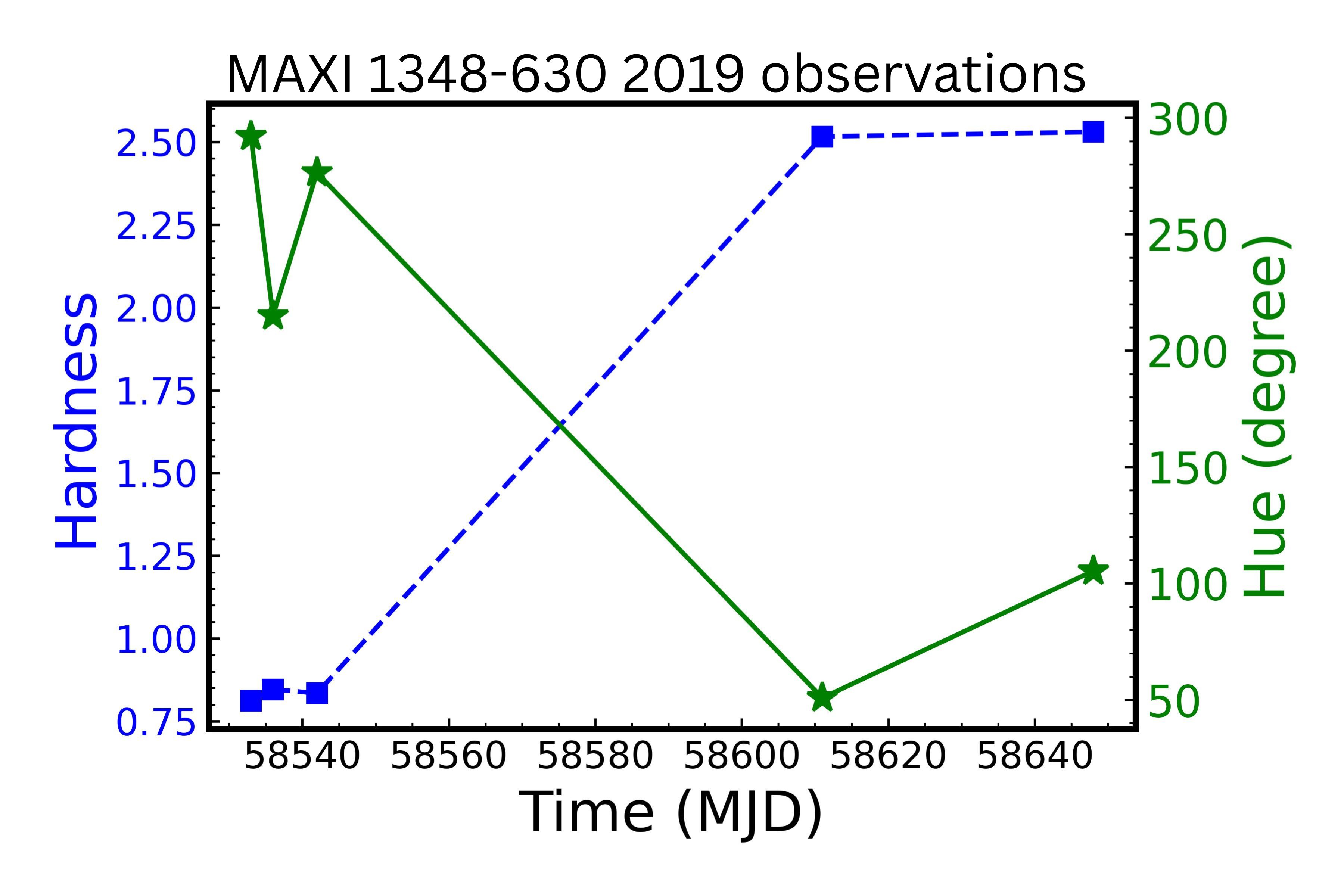}

\caption{Evolution of hue and hardness with time for different data sets in our sample with significant coverage of their states. Green data points represent hue evolution with a star marker. Blue data points represent hardness evolution with a box marker. The name of the source and the year of the outburst are mentioned in the plot title. \label{Fig:3}}
\end{center}
\end{figure*}

\section{Analysis and Results \label{Sec:3}}

\subsection{\textbf{Power-colour analysis} \label{Sec:3.1}}
 
In our analysis, we utilised the clean light curves to produce dead time (42 $\mu$s) corrected Poisson noise-subtracted, Leahy-normalised power spectra \citep{1983ApJ...266..160L},  considering averaged light curve segments of 256 s and a Nyquist frequency of 16 Hz. 
The dead time corrected Poisson power was calculated following the prescription given in \citet{Yadav_2016} for a non-paralyzable detector case, as applicable to LAXPC, based on the formula given in \citet{1995ApJ...449..930Z}. For the following power colour analysis, the power spectra are segmented into four contiguous, logarithmically spaced frequency bands: 0.0039–0.031 Hz, 0.031–0.25 Hz, 0.25–2.0 Hz, and 2.0–16.0 Hz. The variance is calculated by integrating the power for each band. We define two power colour ratios: PC1, representing the variance ratio between the 0.25–2.0 Hz and 0.0039–0.031 Hz bands, and PC2, representing the variance ratio between the 0.031–0.25 Hz and 2.0–16.0 Hz bands, following the prescription given by \citet{2015MNRAS.448.3339H}.
This analysis is performed in the 4–13 keV energy range, where a clear elliptical trajectory emerges in the power-colour diagram, as displayed in the left panel of Figure~\ref {Fig:1}.  
We consider logarithmic scaling of PC1 on the x-axis and PC2 on the y-axis. 
Unique markers are applied for each source, as explained in the legend, to distinguish them in the power wheel. We have used open markers for the low inclination sources like GX 339--4 \citep{2002AJ....123.1741C,2023ApJ...950....5L}, MAXI J1348--630 \citep{2024ApJ...976...61G,2022MNRAS.511.3125J}, and Swift J151857--572147 \citep{2024ApJ...973L...7P}. The data points from the different sources are observed to scatter along this elliptical region. Few data points even show larger scatter from the ellipse due to heightened, peaked broadband as well as narrow peaked power at different frequency regions.
The coordinate of the central point is adopted at (4.51920, 0.453724) to construct the power wheel, following \citet{2015MNRAS.448.3339H}, benefiting from their larger dataset to refine the centre. This calibration ensures the reliability of the result and performs an accurate comparison, even when working with a relatively smaller observational data set, as considered in this work. The semi-major axis of the ellipse is positioned at a 45-degree angle to the positive x- and y-axes, with the clockwise angle from this axis used to represent the hue. The plot is divided into 18 segments, with the hue values grouped in $20^\circ$ intervals and rendered in different colours as mentioned below Figure~\ref{Fig:1}. Segmentation of the power wheel facilitates the identification of different states in BH-LMXBs. 
Following \citet{2015MNRAS.448.3339H}, we identify the spectral states based on the ranges of the hue value, with the hard state, HIMS, SIMS, and soft states spanning 340$^\circ$ to 140$^\circ$, 140$^\circ$ to 220$^\circ$, 220$^\circ$ to 300$^\circ$, and 300$^\circ$ to 20$^\circ$, respectively. It should be noted that these hue ranges identifying the spectral states are taken directly from \citet{2015MNRAS.448.3339H}.  Also, there is an overlap region reported in the hue range of 340$^\circ$-20$^\circ$ (in the clockwise direction) where the hard and the soft states coexist.
Less scatter is observed in the upper part of the power colour wheel when the source is in the harder state, and much higher scatter in the lower part when the source is in the soft state.

The evolution of the RMS-normalised power spectra \citep{2002ApJ...572..392B} for different hue bins is shown in Figure~\ref {Fig:2}, after the removal of the Poisson noise. A single representative power spectrum is selected from each hue bin to provide a clear picture of this evolution. The power spectra are colour-coded in an identical manner to the colour scheme used in the power wheel (Fig.~\ref{Fig:1}) to show hue variation. This visualization presents the systematic power spectral feature transformation that aligns with the source state change, characterised here by the hue parameter.
In the initial hue bins (0$^\circ$- 20$^\circ$, 20$^\circ$- 40$^\circ$, 40$^\circ$- 60$^\circ$, 60$^\circ$- 80$^\circ$), the power remains relatively high, as well as in the last hue bin, which is 340$^\circ$- 360$^\circ$. Then a gradual decline in power is observed up to a hue bin of 140$^\circ$-160$^\circ$. In the hue bin 160$^\circ$-180$^\circ$, a significant drop in the high-frequency power occurs, suggesting the transition from the hard to the HIMS state. 
The power spectrum in the hue bin of 200$^\circ$-220$^\circ$ looks similar to the earlier bin(160$^\circ$-180$^\circ$). Beyond that, the low-frequency power increases as the source goes from HIMS to SIMS state, and a power law-type noise emerges.
This trend continues until the hue bin of 260$^\circ$-280$^\circ$, after which a significant decline in the high-frequency power is observed. Seemingly, a cutoff around 10 Hz appears as the source entered the soft state in the hue range of 280$^\circ$-340$^\circ$.

\begin{figure*}
\centering
\includegraphics[scale=0.33]{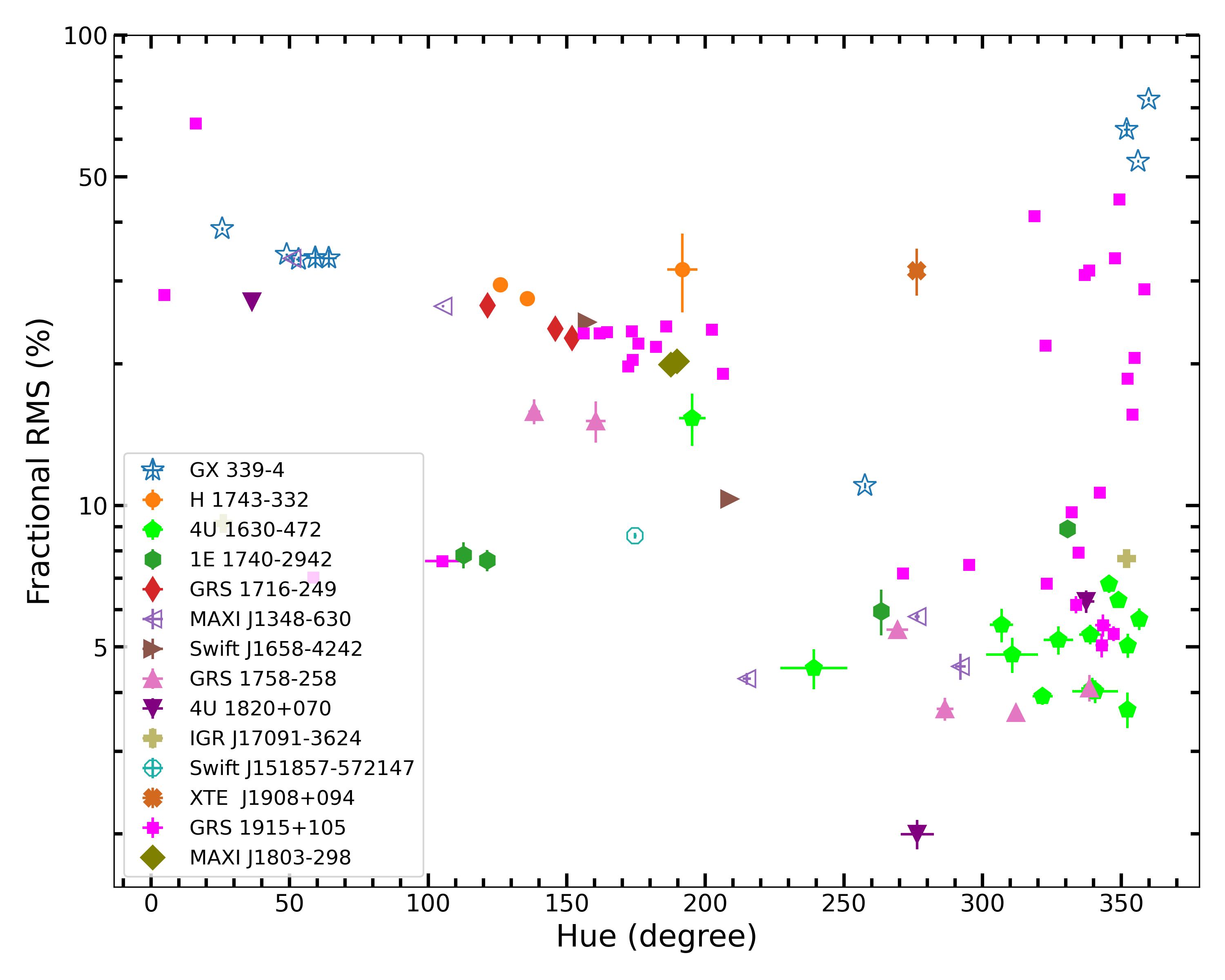}
\includegraphics[scale=0.33]{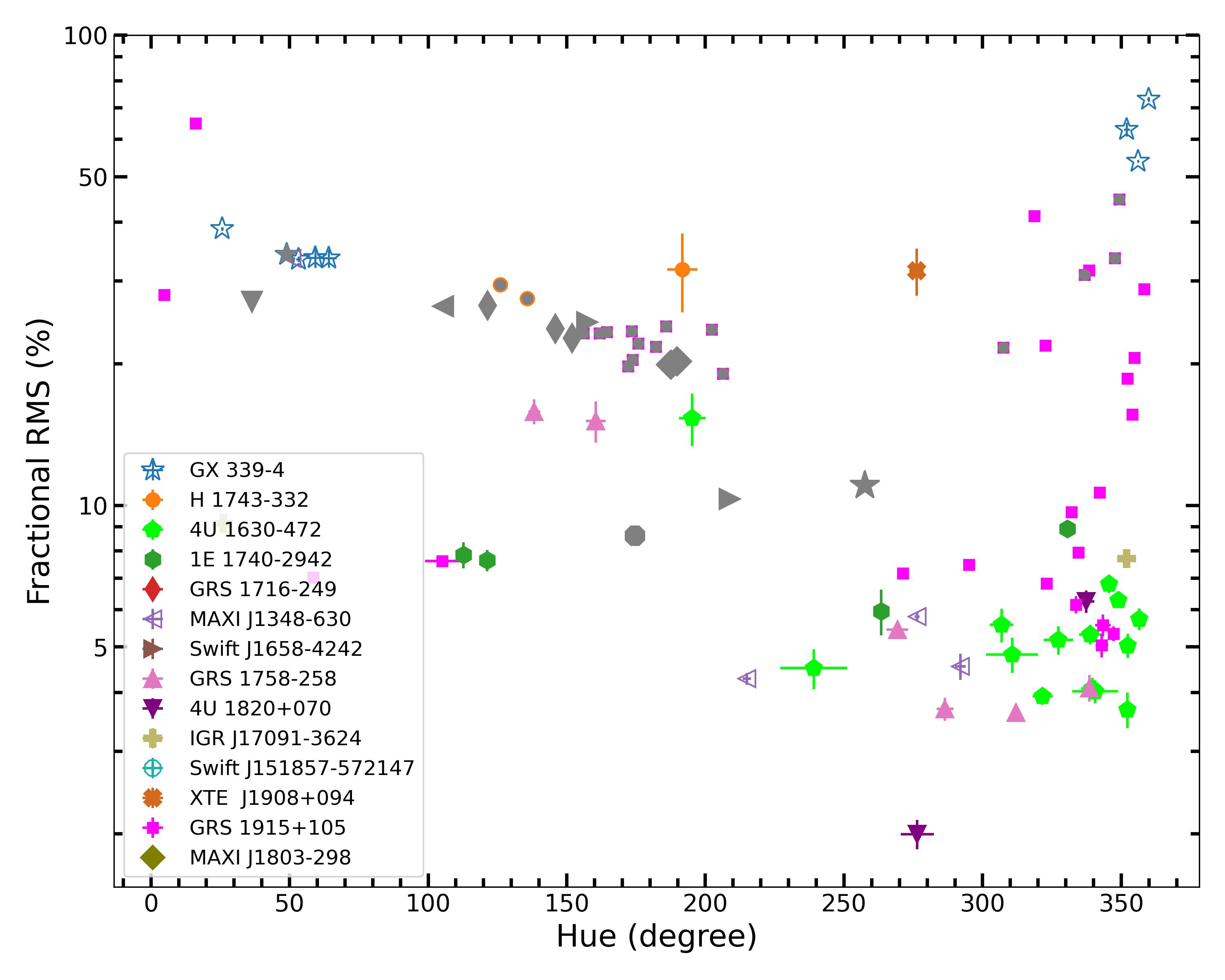}
\caption{Left: Evolution of total fractional RMS amplitude in the total frequency range of 0.0039-16.0 Hz with hue. Different colours and different markers represent different sources as described in the legend.  The open markers represent the observations of the low inclination sources. Note that the y-axis is given in logarithmic units.  Right: The same plot as the left one, with the observations comprising QPOs marked by the grey data points. The solid grey markers represent the high inclination sources with QPO detections, and the open grey markers represent the low inclination sources with QPO detections. \label{Fig:4}}
\end{figure*}
\subsection{\textbf{Hardness and hue evolution with time} \label{Sec:3.2}} 
During outbursts, the source traverses different spectral states, which were typically identified by significant changes in hardness \citep{2004MNRAS.355.1105F}. 
In this work, the evolution of the spectral state is characterised by the change in the hue values.
Hence, for comparison, the evolution of both hardness and hue is analysed over time. Figure~\ref{Fig:3} presents the detailed evolution of hardness and hue during different outburst phases for a representative sample with multiple epoch coverage. 
The 4 to 6 keV energy range is used as the soft band and the 6 to 13 keV range as the hard band, defining hardness as the ratio of count rates between these bands.
The x-axis represents the time in MJD of different sets of the chosen observations to visualise the transitions.
The left y-axis is represented by blue colour, corresponding to the evolution of hardness shown by the box markers, while the right y-axis is represented by green colour, corresponding to the evolution of hue displayed with star markers. 
In certain sources with sufficient data coverage across different phases of outbursts, abrupt shifts are observed in the trend of hardness and hue evolution with time, denoting a certain state transition. Although our data set is limited in some cases, preventing a precise determination of the exact transition epoch, these hardness-hue plots clearly indicate state transitions.
Among the studied cases, this transition behaviour is clearly seen in four cases.
Three correspond to the observations of a high inclination source, GRS 1915+105, in 2016, 2017, and 2018 (each set separated by more than 100 days), and one from MAXI J1348-630,  which is a low inclination source. We separated the observations of GRS 1915+105 according to their observation year, as they correspond to different parts of the source intensity light curve and are separated by larger intervals (more than 100 days).
The source spectral state shows a major transformation when the hardness crosses paths with the hue evolution during this stage.

\begin{figure}

\includegraphics[scale=0.4]{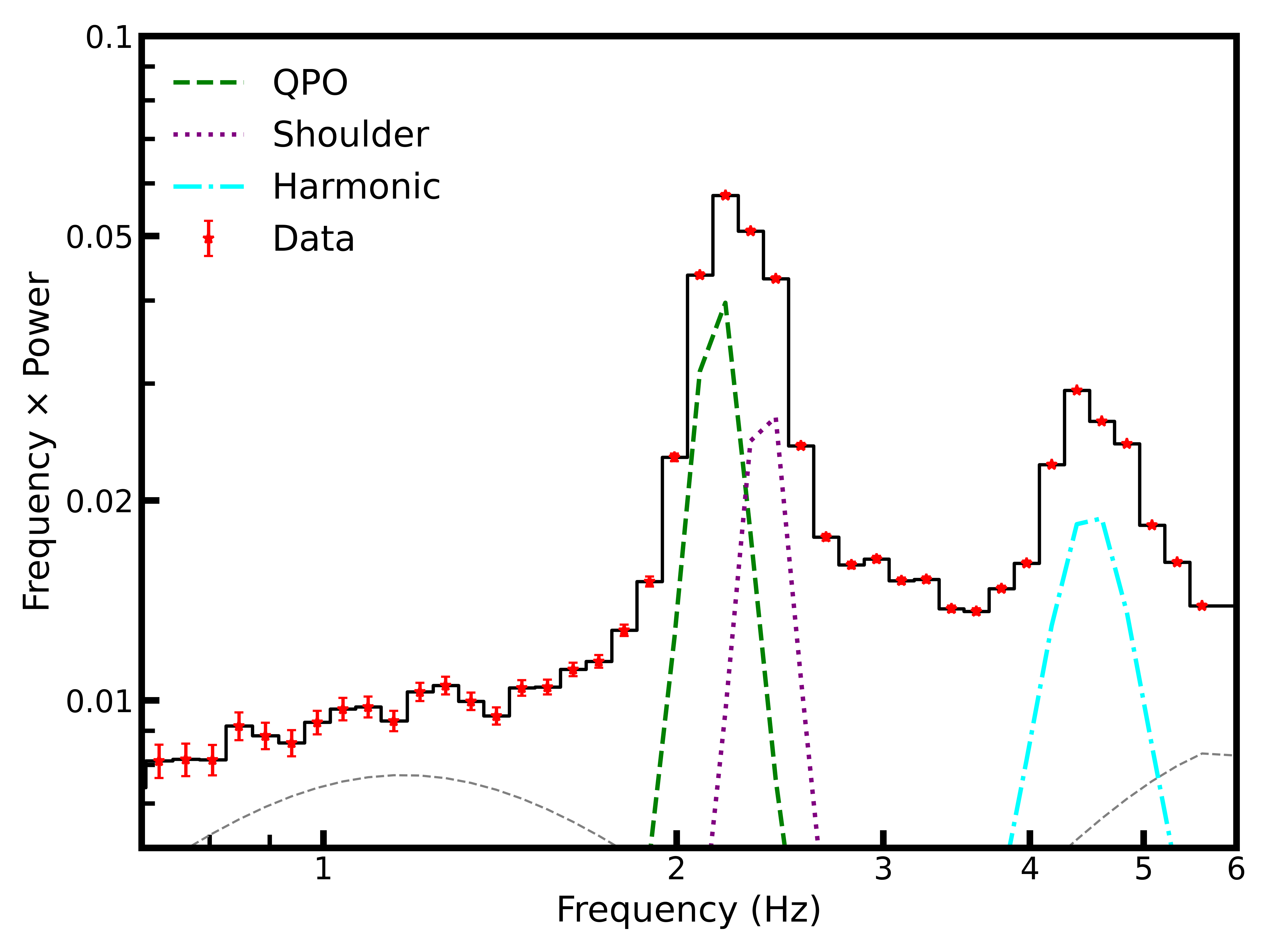}

\caption{Zoomed view of the fitted power spectra of the first observation of the 2019 outburst of GRS 1915+105, focusing on the QPO. 
To fit the QPO peak around 2 Hz (the fundamental QPO), two lorentzians were required. The first lorentzian is referred to as the QPO  (green dashed line), and the second as the shoulder  (purple dotted line). Another peak near $\sim 4.5$ Hz, corresponding to the harmonic, was modelled using one lorentzian (cyan dash-dot line). The individual components are shown using dashed lines of different colours as elaborated in the legend. \label{Fig:5}}
\end{figure}

\subsection{\textbf{Evolution of total RMS of power spectra with hue}\label{Sec:3.3}}

To understand how the total variability of BH-XRBs varies with spectral state evolution, the relationship between RMS amplitude (4-13 keV) against hue is analysed, as shown in the left panel of Figure~\ref{Fig:4}. Different sources are indicated by different markers and different coloured data points for better representation.  The open markers represent the observations of the low inclination sources. The RMS variability is calculated as the square root of the variance of the power spectrum in the 0.0039-16.0 Hz frequency range, and it displays a characteristic ``knee'' pattern in the RMS vs hue evolution plot. A similar trend has been observed using BH-XRB data from \emph{RXTE} (Rossi X-ray Timing Explorer) by \cite{2015MNRAS.448.3339H} and \cite{2015MNRAS.448.3348H}. Specifically, we observe a decline in the RMS as the source transitions from the HIMS to the SIMS state.
Initially, in the hard state (hue $\lesssim$150$^\circ$), the RMS variability remains high, exceeding 30\%  and reaching close to 60 \% in certain cases for sources like GX 339--4, GRS 1915+105,  MAXI J1348--630, and H 1743--332.  All three observations of GRS 1716-249 stay in the range of hue $<$ 150$^\circ$ and RMS $> 20\%$. 
As the hue progresses to the HIMS (140$^\circ$–200$^\circ$, \cite{2015MNRAS.448.3339H}), a gradual decline in the RMS is observed, marking the transition to the softer spectral states. In this range, we observe a higher scatter in the RMS and hue values. The two observations of MAXI J1803-298 fall in the hue range of $< 200$, and the RMS variability is around 20\%. In the hue range of 140$^\circ$ to 200$^\circ$, it is noted that the data points of GRS 1915+105 are clustered in the RMS range of 20-30\%. The observations of GRS 1758-258 and 1E 1740-294 show relatively lower values of RMS in their hue range compared to other observations, indicating comparatively weaker variability of the system in the HIMS state.
The decline of RMS becomes more pronounced above hue values of $200^\circ$, where a sudden jump is observed, aligning with a transition to the SIMS.
Due to limitations of the sampling of our data, this transitional phase is not well-mapped. However, post-transition, the data sampling is sufficiently dense to allow for a critical analysis of hue-RMS evolution in the hue range of 220$^\circ$ to 300$^\circ$.

In the softer state, RMS variability stabilises within a range of approximately 4–8\% for the low inclination sources, and a scatter is observed in the RMS for the highly inclined sources, like GRS 1915+105. 
Most of the data points of 4U 1630-472 fall within the soft region, with RMS values less than 7\% except for one. Two observations of 4U 1820+070 are also in the soft range, and interestingly, one of these observations exhibits the minimum RMS ($\sim$2\%) among all of our observations. 
As the outburst cycle progresses, an increase in RMS is observed, with the source eventually returning to the hard state at the end of the outburst. Around the hue value of 350$^\circ$-360$^\circ$, some sources show higher variability, while some show lower variability. This indicates both hard and soft states, leading to the identification of this region as the overlap region similar to those mentioned by \cite{2015MNRAS.448.3348H}.

\begin{table*}
\begin{center}
\scalebox{0.75}{%
    \begin{tabular}{ccccccccc}
    \hline
    Source & Obs ID &   Hue (degree)& Frequency (Hz) & Width (Hz) & Q factor & Feature$^a$& Inclination\\
   \hline
GRS 1915+105
&9000001124&206.31$\pm$0.57&4.02$\pm$0.02 &0.85$\pm$0.09&4.73$\pm$0.48 &Q\\
&&&4.96$\pm$0.04 &0.95$\pm$0.12&5.20$\pm$0.67 &S\\
&9000001162& 347.63$\pm$0.25&4.86$\pm$0.02 &0.88$\pm$0.06&5.52$\pm$0.38 &Q \\
&&&5.86$\pm$0.09 &1.45$\pm$0.24&4.03$\pm$0.67&S\\
&9000001166&307.44$\pm$0.42&4.49$\pm$0.02 &0.56$\pm$0.10&7.98$\pm$1.46&Q& High\\
&&&5.10$\pm$0.10 &1.39$\pm$0.14 &3.65$\pm$4.88 &S\\
&9000001232&336.75$\pm$0.53&4.85$\pm$0.04 &0.85$\pm$0.12&5.69$\pm$0.82& Q\\
&&&5.81$\pm$0.13 &1.55$\pm$0.31 &3.74$\pm$0.76& S\\
&9000001236&349.22$\pm$0.22&5.71$\pm$0.03 &1.76$\pm$0.11&3.25$\pm$0.20& Q
\\
&9000002000&173.64$\pm$0.28&3.29$\pm$0.04 & 0.64$\pm$0.16&5.18$\pm$1.33& Q\\
&&&3.63$\pm$0.12 &1.19$\pm$0.06&3.05$\pm$0.19& S\\
&9000002006&172.09$\pm$0.28&3.59$\pm$0.01& 0.85$\pm$0.06&4.23$\pm$0.29& Q\\
&&& 4.34$\pm$0.44& 1.92$\pm$0.69&2.26$\pm$0.85& S \\
&&&7.09$\pm$0.09& 3.03$\pm$0.44&2.33$\pm$0.34& H\\
&9000002080&185.77$\pm$0.34&2.33$\pm<$0.01& 0.36$\pm$0.02&6.37$\pm$0.28& Q\\
&&&4.62$\pm$0.02 &1.11$\pm$0.09&4.18 $\pm$0.35&H\\
&9000002110& 182.15$\pm$0.14&3.17$\pm<$0.01& 0.51$\pm$0.04&6.19$\pm$0.43 &Q\\
&&&3.69$\pm$0.03& 1.06$\pm$0.08 &3.48$\pm$0.26&S\\
&&&6.49$\pm$0.07& 2.22$\pm$0.46&2.92$\pm$0.61&H\\
&9000002112&175.71$\pm$0.29&3.28$\pm<$0.01&0.57$\pm$0.03&5.78$\pm$0.29& Q \\
&&&6.53$\pm$0.06& 1.64$\pm$0.24& 3.99$\pm$0.59 &H\\
&9000002190&173.36$\pm$0.40&2.45$\pm<$0.01& 0.30$\pm$0.02 &8.07$\pm$0.62&Q\\
&&&4.93$\pm$0.04& 0.99$\pm$0.14&4.98$\pm$0.69& H\\
&9000002220&164.44$\pm$0.23&1.78$\pm$0.03 &0.04$\pm$0.07&43.35$\pm$77.96 &S \\
&&&1.92$\pm$0.01& 0.19$\pm$0.01&9.78$\pm$0.70& Q\\
&&&3.82$\pm$0.01&0.58$\pm$0.04 &6.55$\pm$0.47 &H\\
&9000002306&156.08$\pm$0.43&1.38$\pm<$0.01 &0.14$\pm<$0.01 &10.19$\pm$0.08 &Q\\
&&&2.82$\pm$0.01 &0.28$\pm$0.03 &10.12$\pm$1.05 &H\\
&9000002334&161.78$\pm$0.45&1.82$\pm<$0.01 &0.17$\pm<$0.01 &10.58$\pm$0.55&Q\\
&&&3.73$\pm$0.03& 0.63$\pm$0.087&5.94$\pm$0.82 &H\\
&9000002812&202.26$\pm$0.41&2.37$\pm<$0.01& 0.23$\pm$0.02&10.46$\pm$1.06& Q \\
&&&2.16$\pm<$0.01& 0.22$\pm$0.02&9.94$\pm$1.05&S \\
&&&4.48$\pm$0.01&0.93$\pm$0.09&4.82$\pm$1.55 &H\\
   \hline
GX 339-4 &9000004218
 &53.28$\pm$0.93 &0.18$\pm<$0.01&0.05$\pm$0.01&3.75$\pm$1.03  & Q& Low\\
 &9000006070&257.59$\pm$0.33& 4.40 $\pm$0.05&0.24 $\pm$0.06&18.43$\pm$4.48& Q& \\
 &&&5.05$\pm$ 0.25&0.59$\pm$0.12&8.44$\pm$1.70&S\\
 &&&9.24$\pm$0.06&1.52$\pm$0.17&6.07$\pm$0.69& H &\\
 \hline
MAXI J1348-630&9000002990 &105.37$\pm$0.35&0.89$\pm<$0.01& 0.25$\pm$0.05&3.62$\pm$0.72&Q& Low
\\\hline
Swift J151857-572147 & 9000006126& 174.66$\pm$0.46&7.93 $\pm$0.04&2.99$\pm$0.12& 2.8$\pm$0.12&Q & Low\\\hline
4U 1820+070 &9000001994&36.43$\pm$0.34&0.05$\pm<$0.01 &0.01$\pm<$ 0.01 &3.85 $\pm$0.94  & Q & High
 \\\hline
H 1743 - 332&9000000364 &135.65 $\pm$0.57&0.60$\pm<$0.01& 0.07$\pm<$0.01&8.49$\pm$2.07& Q& High\\
&&&1.21$\pm<$0.01& 0.10$\pm$0.03&11.84$\pm$3.12&H\\
 &9000001444 & 126.02$\pm$0.61& 0.44$\pm<$0.01 &0.05$\pm<$0.01 &9.49$\pm$0.82& Q\\
 &&&0.88$\pm<$0.01 &0.11$\pm$0.03&8.01$\pm$4.19&H\\
 \hline
GRS 1716-249&9000001034&121.40$\pm$0.35&0.64$\pm<$0.01& 0.07$\pm$0.03&9.01$\pm$3.32 &H  & High\\
&&&0.32$\pm<$0.01& 0.13$\pm$0.02&2.35$\pm$0.45 &Q\\
&9000001140 &151.91$\pm$0.76&
2.52$\pm$0.05& 0.48$\pm$0.21&5.29$\pm$2.33 &H\\
&&&1.20$\pm$0.02 &0.42$\pm$0.09 &2.89$\pm$0.67 &Q
\\
&9000001378 &145.92$\pm$0.61&
1.55$\pm$0.02 &0.63$\pm$0.12&2.45$\pm$0.48 &Q
\\\hline
 Swift J1658.2-4242&9000001910&158.47$\pm$0.33&1.56$\pm$0.01& 0.18$\pm$0.04&8.59$\pm$1.85  & Q& High\\
 &&&1.72$\pm$0.04 &0.38$\pm$0.05&4.53$\pm$1.75&S\\
&&& 3.23$\pm$0.04& 0.71$\pm$0.17&4.57$\pm$2.32&H\\
&9000001940&207.90$\pm$1.21
&6.62$\pm$0.04 &1.83$\pm$0.14&3.62$\pm$0.27 &Q\\\hline

MAXI J1803-298&9000004368&187.55$\pm$0.35&5.39$\pm$0.01& 0.80$\pm$0.04& 6.70$\pm$0.34&Q & High\\
&&&6.46$\pm$0.05&1.07$\pm$0.16&6.04$\pm$0.96&S \\
& 9000004370&189.72$\pm$0.22&5.48$\pm$0.06&0.73$\pm$0.09&7.49$\pm$1.05& Q\\
&&&5.88$\pm$0.06&0.85$\pm$0.08&6.93$\pm$0.71&S\\\hline

                        \end{tabular}}

\begin{flushleft}
\begin{footnotesize}
$^a$ Q: QPO, S: Shoulder, H: harmonic

\end{footnotesize}
\end{flushleft}
\caption{Details of all the detected QPO, harmonic, and shoulders along with the hue value for that particular observation.\label{Tab:2}}
\end{center}
\end{table*}

\begin{figure*}
\begin{center}
\includegraphics[scale=0.5]{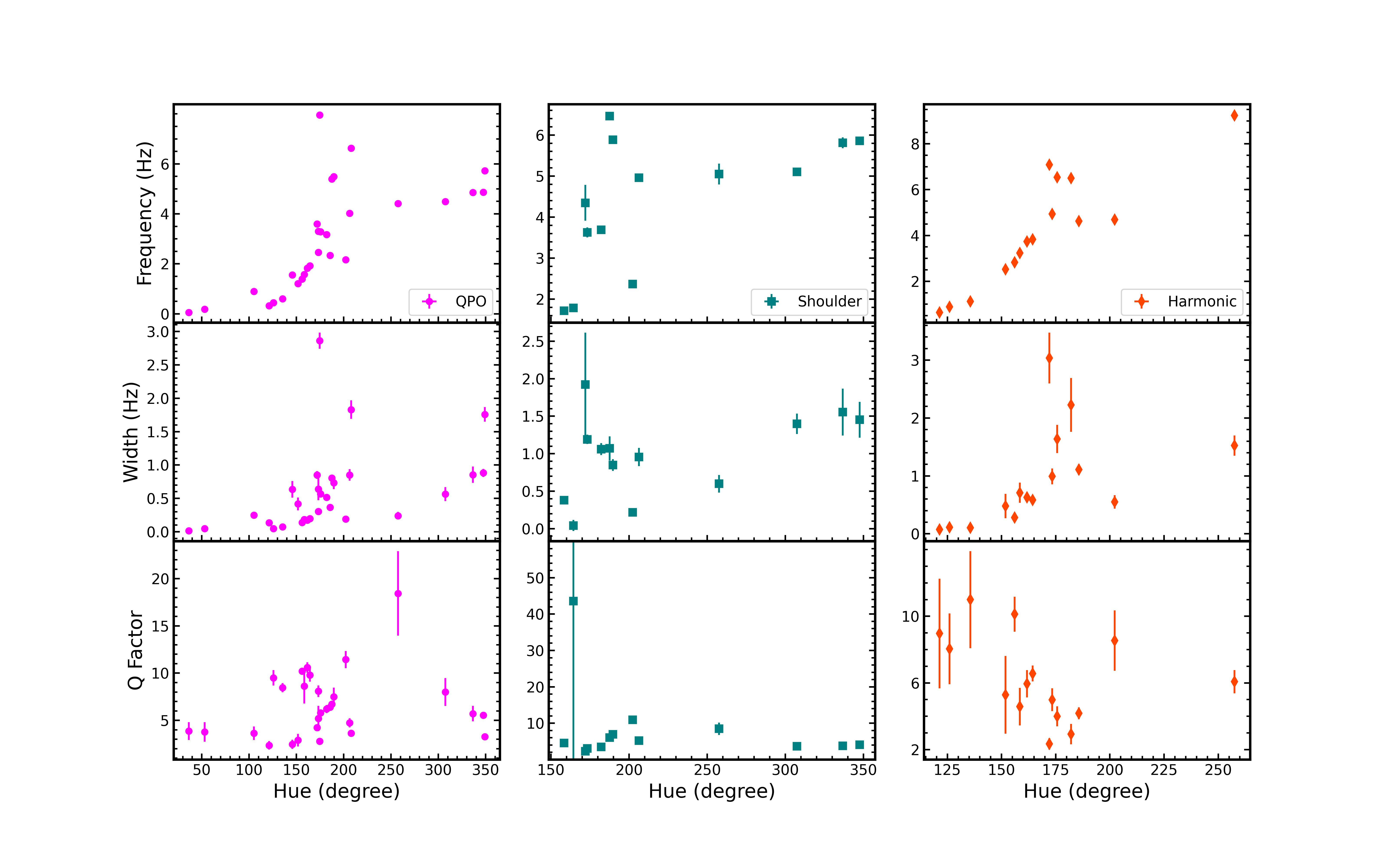}
\vspace*{-1cm}
\caption{Evolution of the central frequency (top row), width (middle row), and $Q$ factor (bottom row) of the detected QPOs represented by the circular marker (left panel), shoulders represented by the square marker (middle panel), and harmonics represented by the diamond markers (right panel) with hue values.  \label{Fig:6}}
\end{center}
\end{figure*}

\subsection{\textbf{Evolution of frequency, width, Q factor of QPO, harmonic and shoulder with hue} \label{Sec:3.4} }

The power spectra are analysed using {\tt XSPEC}, via fitting them using a combination of lorentzians \citep{2002nmgm.meet.2249B}. The lorentzians are added in a sequential manner, and at every step, the addition of the next lorentzian is checked employing an F-test. We put a threshold
of $3\sigma$ significance on the chance probability obtained through the F-test. The addition of a lorentzian is considered only when the chance probability of improvement of the fit on the addition of the lorentzian is less than 0.27\%. In a few cases where the variability is weak, a power law model is used instead, which provides a better fit.
The lorentzian components are characterised by their central frequencies ($f$), widths ($\Delta f$), and quality factors ($Q = f/\Delta f$). The lorentzians are selected with a $Q$ factor greater than 2 to identify the QPOs and other peaked features. 
We thereby identify 29 quasi-periodic oscillations (QPOs) and their associated harmonics in a total of 15 observations and occasional shoulders in 13 of those observations. We did not detect any statistically significant shoulder component along with any of the harmonics.
The grey data points in the power colour diagram in the right panel of Figure~\ref{Fig:1} represent the observations where we detect QPOs. The open grey markers represent the low inclination sources with QPO detection in those observations. A similar representation is used in the right panel of Figure~\ref{Fig:4} to visualise the QPO detections along with the evolution of the RMS with hue. Interestingly, most of the QPO detections fall within the hue range of 100$^\circ$-220$^\circ$, corresponding to the hard and HIMS state region. We have two detections in the hue range of 30$^\circ$-60$^\circ$ associated with a hard state observation of GX 339--4. The four QPO detections in the hue range of $\sim$300$^\circ$-360$^\circ$ correspond to the softer state observations of GRS 1915+105, though it should be noted that the last hue bin corresponds to the overlap region.
We have only one QPO detection in the hue range $\sim$220$^\circ$-300$^\circ$, which is the region of the SIMS state (Figure ~\ref{Fig:1}).

In Figure~\ref{Fig:5}, we present a zoomed-in view of the fitted power spectrum of GRS 1915+105, highlighting the QPO, the shoulder, and the harmonic with different colours and different representative line styles. Two distinct narrow features are clearly visible in the power spectra, representing the QPO fundamental and its harmonic. The QPO feature is well-fitted with two significant lorentzians in some cases. The fundamental QPO peak is shown by the green dashed line, and an additional lorentzian close to the previous one is represented by the purple dotted line. This additional structure, associated with the fundamental QPO peak at a comparatively higher frequency and usually with a lower $Q$ factor, is identified as the shoulder component \citep{1997A&A...322..857B, 2024MNRAS.527.9405M}.
The second narrow feature is fitted with the cyan dash-dot lorentzian, which represents the harmonic of the QPO. Table~\ref{Tab:2} summarises the central frequencies, widths, and $Q$ factors for these QPO, harmonic, and shoulder, along with the hue of that particular observation, while Figure~\ref{Fig:6} illustrates their variation with hue. 
In Figure~\ref{Fig:6}, the left panel represents the results for the QPO fundamentals, the middle panel represents the evolution of the parameters of the shoulders, and the right panel represents the evolution of the parameters for the harmonics. 
Most of the QPOs are detected in the hue range of 100$^\circ$-200$^\circ$ and 300$^\circ$-360$^\circ$. 
Our results demonstrate a clear increasing trend of the QPO frequency with the hue (upper left panel of Figure~\ref {Fig:6}), with a Spearman correlation coefficient of 0.88.
The QPO width also increases with hue, exhibiting a Spearman correlation of 0.72. The $Q$ factor of the QPOs exhibits scattered behaviour throughout the hue range, without a clear monotonic trend.

QPOs are detected across all the accretion states- hard, HIMS, and soft, while shoulders are primarily observed in the HIMS and in the hard state ($\sim$150-350$^\circ$).
As shown in the second column of Figure~\ref {Fig:6}, the shoulders display a scattered frequency distribution in the HIMS ($\sim$150$^\circ$ - 220$^\circ$ ), transitioning to an upward trend in the soft ($\sim$300$^\circ$ - 360$^\circ$) state, with a global Spearman correlation coefficient of 0.64. The width mostly shows scattered evolution with hue, whereas the $Q$ factor exhibits a predominantly stable behaviour. 

The harmonic components exhibit an increasing frequency trend with hue (third column panel of Figure~\ref{Fig:6}), with a Spearman correlation coefficient of 0.89, which is similar to the evolution of the QPO fundamental. Most of the harmonics are detected in the intermediate states with frequencies ranging from $\sim$1-6 Hz. Among them, only one harmonic was observed at 9.24 Hz occurring at a hue value of 257.59$^\circ$.
However, in the 170$^\circ$-190$^\circ$ hue range, harmonic frequencies show more scattered evolution.
The width follows an increasing trend with the frequency, but with a Spearman correlation of 0.76. The $Q$ factor shows a different behaviour. It decreases with hue, with a Spearman correlation coefficient of $-0.48$. 

\begin{figure*}
\begin{center}
\vspace*{-1cm}
\includegraphics[scale=0.6]{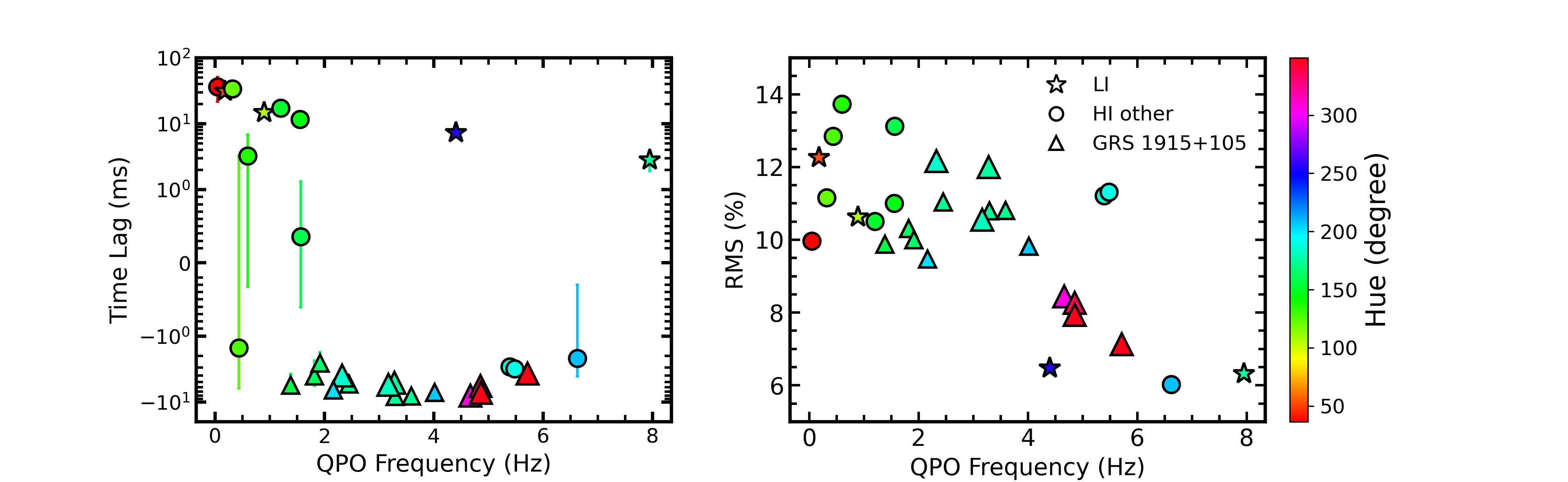}
\caption{Evolution of time lag of the QPO with QPO frequency (left panel) considering 4-10 keV as soft band and 10-30 keV as hard band. High-inclination sources are shown with circular and triangular markers (triangles denote GRS 1915+105), while low-inclination sources are represented by star markers.
Evolution of fractional RMS amplitude (4-30 keV) of the QPO with QPO frequency (right panel). The colours represent the evolution of hue values as elaborated in the colourbar shown on the right.  \label{Fig:7}}

\end{center}
\end{figure*}

\begin{figure*}
\begin{center}
\includegraphics[scale=0.48]{combined_rms_energy_inclination2.jpg}
\caption{ Evolution of energy-dependent RMS  for high inclination (upper panel) and across four low inclination (lower panel) QPO observations to highlight changes in RMS spectral trends with different hue and QPO central frequency ($f$) values. \label{Fig:8}}

\includegraphics[scale=0.55]{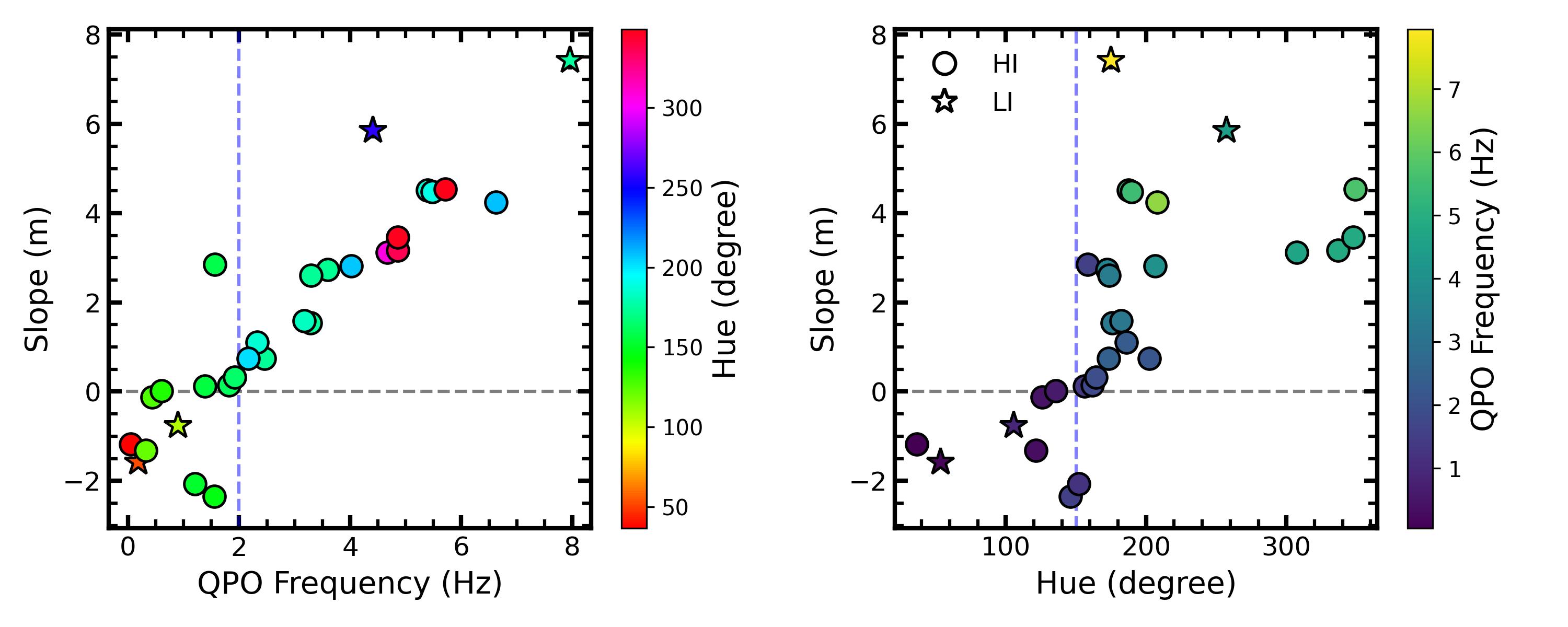}
\caption{ Evolution of the slope ($m$) of the fitted logarithmic function to the RMS spectra with QPO frequency (left panel) and hue (right panel).  Circular and star markers denote high- and low-inclination sources, respectively. In the left panel, different colours represent different hue values (as shown in the corresponding colour bar), and in the right panel, different colours represent values of QPO frequency (as shown in the corresponding colour bar).  \label{Fig:9}}
\end{center}
\end{figure*}
\begin{figure*}
\begin{center}
\includegraphics[scale=0.48]{combined_lag_energy_inclination.jpg}
\caption{Evolution of energy-dependent time lag for high inclination (upper panel) and across four low inclination (lower panel) QPO observations to highlight changes in RMS spectral trends with different hue and QPO central frequency ($f$) values. Here, we have excluded the reference band of 4.0-5.0 keV during fitting.
\label{Fig:10}}

\includegraphics[scale=0.55]{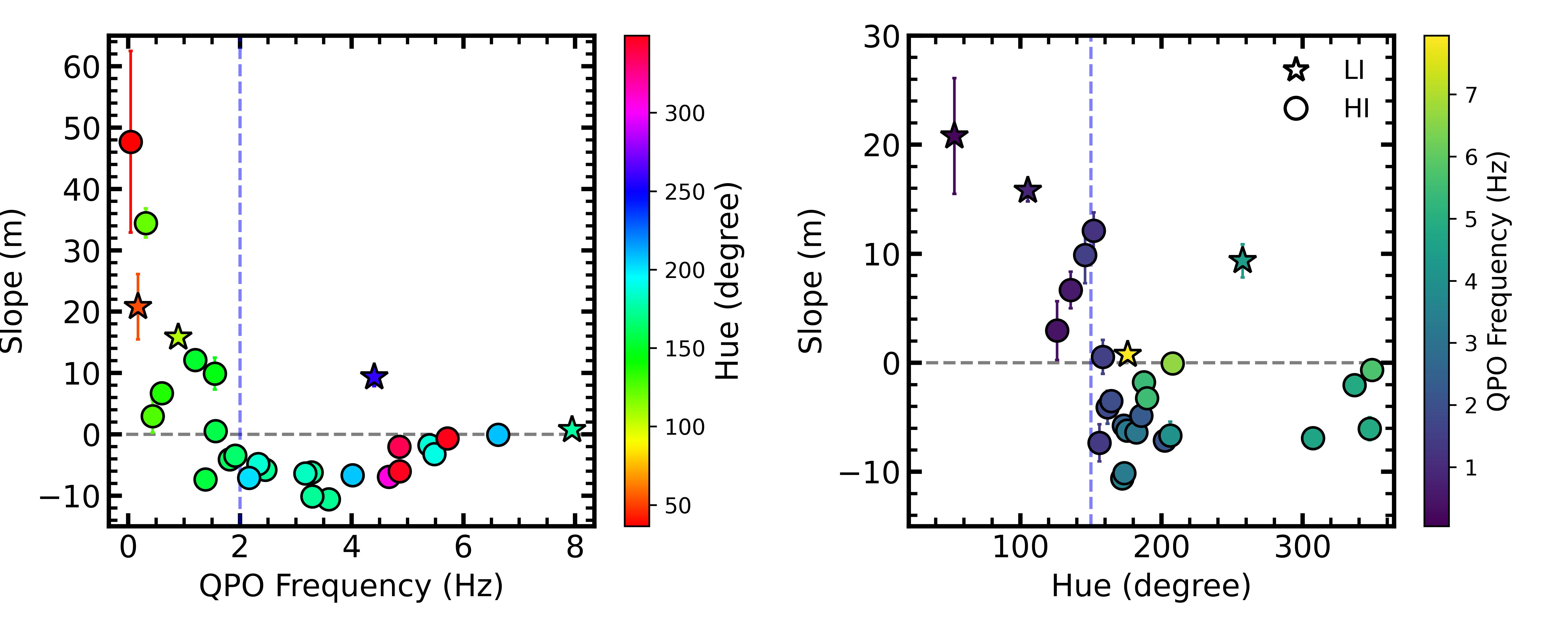}
\caption{Evolution of the slope ($m$) of the fitted logarithmic function to the time lag spectra with QPO frequency (left panel) and hue (right panel) for all sources.  Circular and star markers denote high- and low-inclination sources, respectively. In the left panel, different colours represent different hue values, and in the right panel, different colours represent values of QPO frequency. The dashed horizontal line in both plots represents the zero slope. The dashed vertical line in the left plot represents the QPO frequency around which we see the transition of the slope, and the dashed vertical line in the right panel represents the hue value where we see the transition.  \label{Fig:11}}
\end{center}
\end{figure*}

\subsection{\textbf{Average RMS and time lag evolution  \label{Sec:3.5}}}
We also examined how the energy-integrated RMS variability strength and time lag vary with the states of the source. The RMS-normalised power spectra are computed in 4–30 keV energy bands by using the {\tt laxpc\_find\_freqlag} tool in the frequency range of 0.0039–16 Hz.  
The RMS variability amplitude of QPOs is estimated using the sum of the RMS-normalised power spectral density over the frequency band from ($f_{QPO} -\Delta f/2$) to ($f_{QPO} +\Delta f/2$), where $f_{QPO}$ denotes the QPO centroid frequency and $\Delta f$ is the width of the QPO.
We chose the summation method to maintain consistency with our time-lag analysis discussed in the next paragraph, where we similarly averaged the time lag over each frequency bin within the QPO frequency range.
We further calculate the average time lag for the QPOs by considering the 4-10 keV energy band as the soft band and 10-30 keV as the hard band.
To calculate the time lag, we use the method mentioned in \citet{1999ApJ...510..874N}.
A frequency resolution of 0.0039 Hz is considered for generating the cross-spectra, and consequently, time lags are obtained from each 0.0039 Hz interval of the cross-spectra.
Finally, we average the time lags in the QPO frequency range to calculate the time lag of the QPO. This approach ensures that the measurement specifically focuses on the lag associated with the QPO feature, accurately capturing its intrinsic contribution within the defined frequency range, between the FWHM of the QPO. The errors are calculated on the averaged time lag of the QPO by propagating the errors of the individual lag values. The majority (25 out of 29) of the QPO detections in our sample are associated with high-inclination sources, with only four QPOs associated with low inclination systems, as summarised in Table~\ref{Tab:2}. 
The right panel of Figure~\ref{Fig:7} shows the variation of the averaged QPO RMS with the QPO frequency. High-inclination sources are shown with circular and triangular markers, with the latter specifically highlighting GRS 1915+105, which contributes 15 out of the 25 detected QPOs in this group. The star markers represent the low inclination sources. The data points are colour-coded according to the hue values of the corresponding observations, as indicated by the colour bar. At lower frequencies, the RMS of the QPO is scattered around 10-14 \% till the QPO frequency of $\sim$4 Hz. Then, a decreasing trend of the QPO RMS is observed as the frequency increases, with the RMS as $\sim$6\% as the frequency crosses 6 Hz.  
The two QPOs detected around a frequency of 5.5 Hz, showing a sudden increase of RMS compared to the neighbouring QPOS, belong to the source MAXI J1803-298.
The RMS amplitudes for low inclination sources decreased steadily with frequency from $\sim$12\% to $\sim$6\%.

The QPO time lag for all sources is plotted against the QPO central frequency in the left panel of Figure~\ref{Fig:7}.  
For the high inclination sources, we observe hard lags of $\sim$36 ms for the lowest frequency (0.05 Hz).  The time lag appears to be decreasing as the frequency increases up to approximately 1.5 Hz. However, this trend cannot be claimed to be significant due to the large uncertainties in the QPO time lag for some observations. The four circular data points below 2 Hz, which exhibit time lags above 10 ms and show a decreasing trend with QPO frequency, are from three different sources: one from 4U 1820+070, two data points are from H 1743 - 332, and another data point is from GRS 1716-249 (Table ~\ref{Tab:2}). Although these data points show a similar trend, it is not indicative of any distinct source-specific behavior.
A distinct transition of the time lag from hard lag to soft lag occurs for the high inclination sources around the QPO frequency range of 2 Hz. The time lags remain in the soft lag region with increasing frequency after the transition. We observe that the transition of time lag is associated with a hue value of $\sim$150$^\circ$. 
In Figure~\ref{Fig:7}, we can see that the triangular-marked data points, representing the GRS 1915+105 QPOs, exclusively show soft lags. This indicates that, considering just the GRS 1915+105 source, the transition of the QPO time lags might not have been detected. 
This plot also represents one of the advantages of this entire work, considering that a comprehensive sample of suitable observations with significant QPO detections is ideal for performing the time-lag analysis.

The four observations of the low inclination sources represented by the star markers in Figure~\ref{Fig:7} exhibit exclusively positive lags for both low ($<$ 2 Hz) and high ( $>$ 2 Hz) frequency QPOs. 
The time lag steadily decreased as the QPO frequency increased for low inclination sources. We observe hard lags of $\sim$31 ms for the lowest frequency (0.18 Hz) and a hard lag of nearly 3 ms for the highest frequency (7.95 Hz). The time lag of low inclination sources at lower frequencies ($<2$ Hz) exhibits a similar trend to that of high-inclination sources, whereas at higher frequencies they remain positive as opposed to the sign reversal of the lag in high inclination sources.

\begin{figure*}
\begin{center}
\includegraphics[scale=0.56]{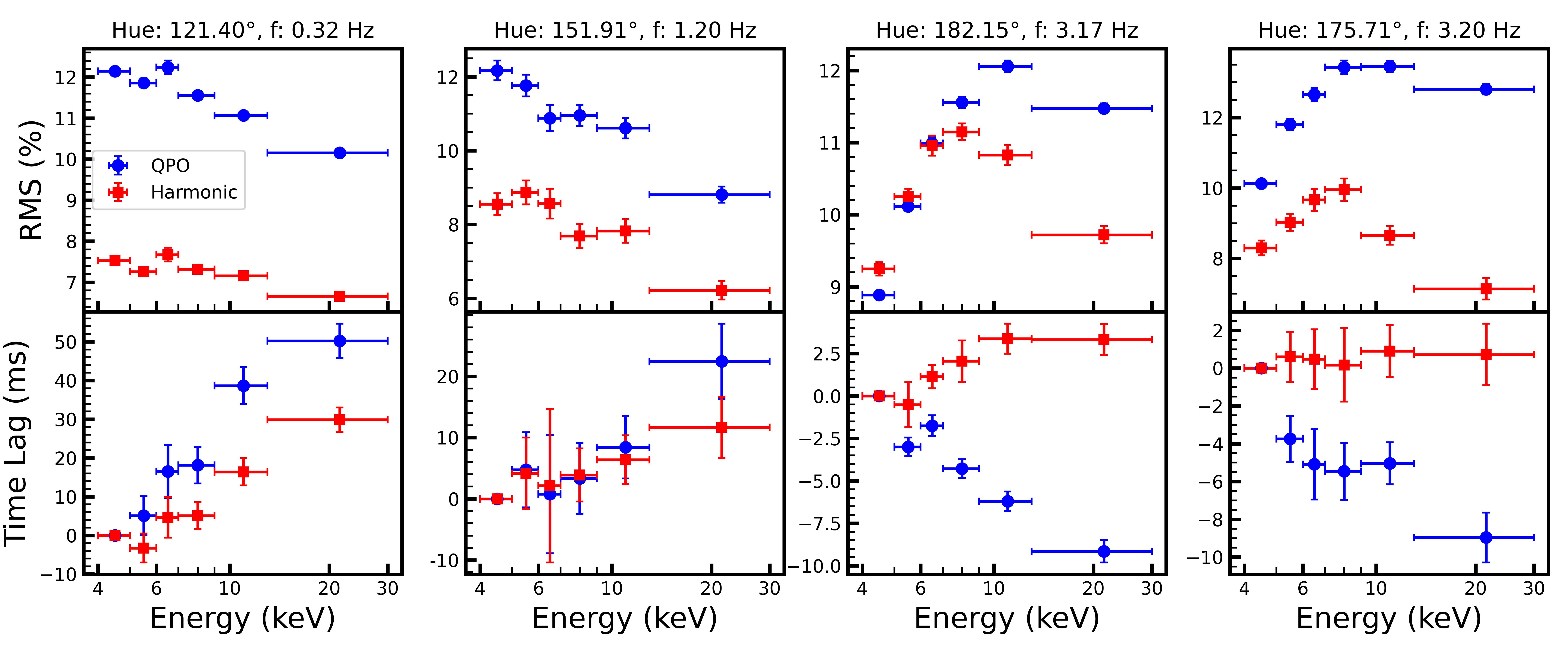}
\caption{The energy dependence of the RMS amplitude and time lag for the QPO and its harmonic is presented for the high inclination sources, with blue circle and red square symbols, respectively. Four representative observations are shown to highlight the diversity in their evolution across different hue values and QPO frequencies. The hue and the corresponding QPO frequency ($f$) for each case are indicated above the respective panels. \label{Fig:12}}
\end{center}
\end{figure*}

\subsection{\textbf{Energy dependency of RMS of QPO  \label{Sec:3.6}}}
To investigate the energy-dependent variability properties of the QPO, the 4.0-30.0 keV energy range is divided into 6 segments (4.0-5.0 keV, 5.0–6.0 keV, 6.0–7.0 keV, 7.0–9.0 keV, 9.0–13.0 keV, and 13.0–30.0 keV). To eliminate the effect of background contamination, data above 30 keV is excluded, as background dominates the intensity in this energy range. Our analysis shows evidence for systematic changes in the energy evolution of both the RMS and time lags with the QPO frequency. 
For the RMS analysis, we identify the lorentzian component corresponding to the QPO from the multi-lorentzian fit of the power density spectrum.
We then calculate the sum of the RMS-normalised power spectral density for different energy ranges over the frequency band from ($f_{QPO} -\Delta f/2$) to ($f_{QPO} +\Delta f/2$), taking $f_{QPO}$ and $\Delta f$ as mentioned in Table~\ref{Tab:2}.

To quantify the increasing and decreasing trend of the RMS spectra, we use a logarithmic function to fit the RMS spectra \citep{2020MNRAS.494.1375Z}, 
$$f(E) = m \times \log\left(\frac{E}{E_0}\right) + b$$
where $m$ is the slope of the RMS vs. the energy relationship, $E_0$ is the lowest energy level, and $b$ is a constant.
It is important to emphasize that this fitting is solely for the purpose of quantifying the increasing or decreasing trend, and not for modelling the RMS spectra.

The upper panel of Figure~\ref {Fig:8} represents four of the RMS spectra, selected to show the evolution of their trend with the QPO frequency and hue of the high inclination sources in our sample.
The different colours here represent the different hue ranges, consistent with our earlier hue colour plots (Figure~\ref {Fig:1}). The evolution of the RMS energy spectra for the four QPO observations of the low inclination sources is shown in the lower panels of Figure~\ref{Fig:8}.

All the RMS energy spectra were fitted with the logarithmic function mentioned earlier. In Figure~\ref {Fig:9}, the circular data points show the variation of the slope $m$ of the fitted model with the QPO frequency (left panel) and the hue (right panel) for the high inclination sources, and the star markers represent the low inclination sources. 
Before $\sim 2$ Hz, the slope shows mostly negative values (left plot of Figure~\ref{Fig:9}), whereas post this frequency, the slope exhibits exclusively positive values. The one data point which shows a positive slope below 2 Hz range is from the first observation of the  source Swift J1658.2-4242.
Beyond $\sim$2 Hz, the slope of the RMS spectra, which becomes exclusively positive, increases with the QPO frequency.
Therefore, there is an indication of a transition of the slope, which is more prominent in its evolution with the hue (right panel of Figure~\ref {Fig:9}). 
Here, the transition is evident in both plots, but the slope vs. hue plot shows the transition region is similar (around a hue value of 150$^\circ$) for both the high and low inclination sources, though for the low inclination sources, we can not claim it independently due to fewer data points. The QPO of 4.70 Hz detected in GX~339$-$4, also reported in \cite{10.1093/mnras/staf2280} where it was marked as O4$^{*}$, also shows an increasing trend like the ones we observed for high ($>$ 2) frequency QPOs.

\subsection{\textbf{Energy dependency of time lag of QPO } \label{Sec:3.7}}
For the energy-dependent time lag analysis of the QPO, the lowest energy band, 4.0–5.0 keV, is chosen as the reference band, and time lags for other energy bands are determined accordingly. All 29 time lag spectra of the QPOs are fitted with a similar logarithmic model mentioned in Section~\ref{Sec:3.6}, in order to estimate the slope on the time lag energy spectra. 
In the representative 4 lag spectra of the high inclination sources, an increasing trend is observed for hue of 36.43$^\circ$ and 151.91$^\circ$ as seen in the first two subplots in the upper panels of Figure~\ref {Fig:10}. The next two subplots in the upper panels show a decreasing trend in the lag spectra. The four lag spectra from the low inclination sources covering a wide range of hue values (53.28$^\circ$-257.59$^\circ$) shown in the lower panels of Figure~\ref{Fig:10}, exhibit a clear increasing trend with energy. In the left panel of Figure~\ref{Fig:11}, we plot the slope of the time lag spectra as a function of QPO frequency for all sources, where high-inclination sources are represented by circular data points and low-inclination sources by star markers. The colours here represent the evolution of the hue. We note that at QPO frequencies below 2 Hz, the slope of lag spectra is mostly positive, except for three observations. whereas beyond 2 Hz, this is positive for the high inclination sources.
In the right panel of Figure~\ref {Fig:11}, we plot the slope with the hue and observe that the transition of the slope occurs again near the hue value of 150$^\circ$ for the high inclination sources. The colours here represent the evolution of the QPO frequency, as displayed by the colour bar. Similar to the RMS slope analysis, the transition is clearer in the slope vs. hue plot.
In the evolution of the slope of the lag spectra, in particular, the sign reversal further emphasises the hue as a more reliable and sensitive parameter for capturing state transitions.
The slope of the time lag spectra of the low inclination sources decreases with QPO frequency (left panel of Figure~\ref {Fig:11})

\subsection{Difference between the nature of the QPO and the harmonic \label{Sec:3.8}}
We identify significant detections of harmonic features in the power spectra for 15 observations in our sample.  Only one of the harmonics is associated with a low inclination source (GX 339--4), making it difficult to draw strong conclusions about its dependence on the QPO frequency and spectral state for the low inclination case. All the other harmonics detected in our analysis correspond to the high inclination sources. Using the methodology described in Sections ~\ref{Sec:3.6} and ~\ref{Sec:3.7}, we calculate the RMS spectra and the time lag spectra of the harmonics and compare them against similar properties of the corresponding QPOs.  The harmonic from the low inclination source behaves similarly to the corresponding fundamental QPO (at 4.40 Hz). The RMS and time lag spectra of both the QPO and its harmonic showed an increasing trend with energy. In contrast, the behaviour observed in the high-inclination sources shows an intriguing pattern. The comparative representation of the RMS spectra and the time lag spectra of the QPO (blue circle markers) and the harmonics (red box markers) of the high inclination sources is depicted in Figure~\ref {Fig:12} for different values of hue and QPO (fundamental) frequency.
The RMS spectra show similar trends for both QPOs and harmonics across different hue ranges and QPO frequencies, with the harmonic exhibiting lower amplitudes of variability, as shown by the panels in the upper row of Figure \ref{Fig:12}. However, the time lag spectra reveal intriguing differences in their evolution.
In the first panel, in the bottom row of Figure~\ref {Fig:12}, the time lag of the QPO increases with energy, reaching up to 50 ms in the highest energy band. The harmonic lag also increases with energy, reaching a maximum value of $\sim$30 ms in the last energy band.
In the third panel of the bottom row, the time lag spectra of the QPO and harmonic exhibit opposing trends.
The QPO time lag decreases with energy, transitioning to a soft lag of approximately -9 ms at the highest energy range. In contrast, the harmonic time lag does not follow this trend; instead, it increases with energy. 
Interestingly, the magnitude of the harmonic lag at the highest energy band is $\sim$3 ms, which is much less than that of the QPO lag. 
In the fourth panel, the harmonic lag remains close to zero, considering the error bars throughout the energy band. 
These observations suggest that, although the RMS evolution of QPOs and harmonics is similar, their time lag spectra exhibit distinct behaviours, particularly as the hue increases.

\section{Discussion} \label{Sec:4}
\subsection{Hue as state identifier \label{Sec:4.1}}
In this work, we performed the power-colour analysis using the high-timing resolution of \emph{AstroSat} for the very first time to generate a state-dependent characterisation of variability in BH-XRBs. The power-colour diagram offers an efficient and measurable technique for the quick identification of the different accretion states. This method also provides a valuable framework for placing different sources onto a common diagram, allowing for meaningful comparisons of their variability characteristics, which is challenging using traditional HID. It also helps to reduce instrument-related dependencies, making it a versatile tool.
In the hard state, the sources are compactly clustered towards the upper part of the power wheel as illustrated in Figure~\ref{Fig:1}, indicating relatively stable variability. In contrast, the sources in the soft state are scattered, implying that the accretion variability is more complex. Figure~\ref{Fig:2} illustrates the change in the variability pattern as the hue evolves. 
The power spectra show subtle variation across the second to eighth hue bin corresponding to the hard state region ($20^\circ$-$160^\circ$). In the subsequent hue bins ($160^\circ$-$220^\circ$), a decline in high-frequency variability is observed, consistent with the diminishing contribution from the corona \citep{2006ARA&A..44...49R,2010MNRAS.403...61D}, as the source transits from the harder to the softer state.
After hue value of $220^\circ$, we observe a significant drop in the total RMS (Figure ~\ref{Fig:4}), indicating the HIMS to SIMS transition. Further changes in the SIMS power spectra are evident after the hue value of $280^\circ$, including a cutoff at 10 Hz, which indicates the transition to the soft state.
Interestingly, in the first and final hue bin ($0^\circ$-$20^\circ$ and $340^\circ$-$360^\circ$), we observe the power spectra with soft and hard state characteristics, indicating the overlap region previously reported by \citet{2015MNRAS.448.3339H}, where the hard and the soft states coexist. 
GRS 1915+105 shows diverse variability classes in its light curve, and we examined the effect of these different variability behaviours on the hue estimation. We have adopted the classification of the state of variability from previously reported works \citep{2022MNRAS.510.3019A, 2022MNRAS.512.2508M, 2024MNRAS.527.4739M}, and we checked the corresponding variability classes against the hue values. For GRS 1915+105, six major variability classes are identified corresponding to our sample ($\delta, ~\omega, ~\beta,  ~\theta,  ~\rho$, and $\chi$). Most of these classes are concentrated within the hue range of 250$^\circ$–360$^\circ$. The $\rho$ class observations are tightly grouped between hue 300$^\circ$–350$^\circ$. Notably, all the $\chi$  class observations are confined to the hue interval of 150$^\circ$–220$^\circ$, corresponding to the HIMS region, where no other variability classes are detected. Distinctly different power spectral behaviour has previously been reported for this class, compared to the other 5 classes \citep{2022MNRAS.512.2508M} in consistency with our hue results. Except for the $\chi$ class, we did not observe any systematic effect of the variability classes on hue estimation. The flaring activity reported in the light curves at MJD 57892.74 and MJD 57996 \citep{2022MNRAS.512.2508M,2025ApJ...984..118D} falls in the variability class of $\rho$ and $\omega$ variability class \citep{2022MNRAS.510.3019A} and the corresponding hue values for these two observations are 349$^\circ$ and 358$^\circ$, respectively.

Most of the QPOs present in our sample are detected in the hue range of 20$^\circ$-220$^\circ$ (right panel of Figure ~\ref{Fig:1}), which corresponds to the hard and HIMS states where QPOs are typically observed. However, four QPOs from GRS 1915+105 are detected within the hue range of 300$^\circ$-360$^\circ$. It is important to note that the hue range from 340$^\circ$-20$^\circ$ (clockwise) represents an overlap zone exhibiting features of both hard and soft states. 
Overall, no distinct behaviour of these QPOs is observed compared to the QPOs detected in the relatively harder state.
QPOs in the softer state have been reported earlier as well in Cygnus X-1, Swift J1727.8--1613, and 4U 1630-47 \citep{2021ApJ...919...46Y, 2025arXiv250507938B, 2025arXiv251101883P}. 
These four QPO detections of GRS 1915+105 near the overlap region correspond to the $\rho$ class, whereas all the other QPO observations of this source reported in Table~\ref {Tab:2} fall in the $\chi$ class. Also, our RMS and time lag analysis of QPOs show that near hue value of $150^\circ$, there is a significant transition which corresponds to the hard to HIMS transition.
Thus, based on our analysis, we classify the hue range as follows: Hard state ($20^\circ$-$150^\circ$), HIMS ($150^\circ$-$220^\circ$), SIMS ($220^\circ$-$280^\circ$), soft ($280^\circ$-$340^\circ$), and overlap region ($340^\circ$-$20^\circ$).
Due to the high timing sensitivity of \emph{AstroSat}, we were also able to pinpoint the hue value corresponding to the hard to HIMS transition. While earlier studies reported this transition around a hue value of $140^\circ$, our combined RMS and time-lag analysis suggests a revised transition value near $150^\circ$.

The drop in RMS above hue 200$^\circ$ is likely linked to the transition to SIMS, indicating changes in accretion properties and coronal geometry. This behaviour is in accordance with previous findings of \citet{2015MNRAS.448.3339H, 2015MNRAS.448.3348H, 2018MNRAS.481.3761G}, that relate the dampening of the low-frequency variability to alterations in disk geometry and turbulence after the hue value of $\sim$ $220^\circ$. The lack of abundant data from \emph{AstroSat} during state transitions hampers a detailed study of the ``knee'' in the RMS-hue diagram. 
In our analysis, we have noticed that the source GRS 1758-258 and 1E 1740-294 show reduced variability compared to neighbouring sources (Figure~\ref{Fig:4}), leading to a more scattered distribution of data points in the HIMS region. One possible explanation could be that, since GRS 1758-258 and 1E 1740-294 are situated close to the densely populated field of the Galactic centre \citep{1991A&A...247L..29S,1999ApJ...525..901M}, the lesser variability could be attributed to contamination of the emission from other X-ray sources within the wide field of view of LAXPC ($1^\circ \times 1^\circ$). 
We further examined the inclination dependency of the power colour diagram, but could not find any significant change. We also investigated the effect of the presence of the QPO power on the hue vs RMS evolution as suggested by \cite{2015MNRAS.448.3348H}. We did not observe a clear distinction between the low and high inclination sources.
But, at hue values close to 360$^\circ$, GX 339–4 (a low inclination source) displays enhanced RMS variability, which is similar to the trend reported by \citet{2015MNRAS.448.3348H}, where low inclination systems were found to exhibit comparatively higher RMS variability.
Our results in section ~\ref{Sec:3.2} indicate that the monitoring of hardness with hue in BH-XRBs can be used as a state transition indicator (Figure ~\ref{Fig:3}). This identification capability is useful for scheduling future observations and theoretical modelling.  
The variability evolution of BH-XRBs across state transition is supported by our measurements of sudden transitions in hardness and hue during outbursts in sources like GRS 1915+105 and MAXI J1348-630.

\subsection{\textbf{Evolution of the QPO properties}  \label{Sec:4.2}}
We have detected 29 QPOs in our sample of BH-XRBs and examined the averaged RMS, time lag behaviour, along with the slope of the energy spectra of RMS and time lag of the QPO. 
In the case of the  4 QPOs from low inclination sources, as the QPO frequency increases, the lags remain hard, but their magnitude decreases (left panel of Figure~\ref{Fig:7}). The energy-dependent lag spectrum also shows only a positive trend for the low inclination sources (lower panel of Figure~\ref{Fig:10}).
Although our sample of low-inclination sources is limited, similar lag evolution has been reported previously for such sources by \citet{2020MNRAS.497.4222C, 2022MNRAS.514.2839A, 2017ApJ...845..143Z}.
We also examined the QPO at 4.70 Hz detected in GX~339$-$4, recently reported in \cite{10.1093/mnras/staf2280}.  
However, this observation is not included in our final analysis, as it does not satisfy our sample selection criteria outlined in section 2.
For this QPO, the averaged RMS ($6.67 \pm 0.06\%$) and averaged lag ($9.22 \pm 0.69$ ms) values, along with the energy-dependent behaviour of these parameters, follow trends similar to those observed in  Figure~\ref{Fig:7} for the low inclination sources, further substantiating our findings for this type of source.
For high-inclination sources, the time lag shows a transition from hard to soft lags at frequencies near $\sim$2 Hz.
We find that the lag behaviour exhibits a deviation at higher frequencies (beyond $\sim$2 Hz) from that previously reported by \citet{2020MNRAS.494.1375Z}. 
Though at lower frequencies (below 2 Hz), the lag showed similar evolution, whereas beyond $\sim$2 Hz, the lags remain confined within $\lesssim$10 ms (left panel of Figure~\ref{Fig:7}). 
Correlating the transition of the slope of the lag spectra with the hue value provided strong evidence that the QPO lag transition (Figure~\ref{Fig:11}) around 2 Hz corresponds to the hard to HIMS state transition region (occurring at a hue value of $\sim 150^\circ$).
The transition of the lag spectra from positive to negative trend has been shown earlier using \emph{RXTE} data near a QPO frequency of 2 Hz for GRS 1915+105 by \citep{2016MNRAS.458.3655V,2020MNRAS.494.1375Z}.
Several theoretical models have been proposed earlier to explain the time lag transition involving processes, like photon interaction delays \citep{2000ApJ...541..883R}, Compton down-scattering of these hard photons \citep{2000ApJ...538L.137N}, and switching of the spectral shape of inhomogeneous inner flow \citep{2016MNRAS.458.3655V}.
However, these models fall short as there is no clear observational evidence of such a spectral change or an interaction delay of the required magnitude.
All of these results provided certain insights, but significant ambiguities remain regarding the true cause of the lag transition.

This distinct change in lag behaviour across the transition offers a fresh insight into the evolving coronal geometry.
Considering the simplistic picture involving a reverberation scenario, the maximum observed hard lag of 36 ms (Figure ~\ref{Fig:7}) implies a light travel distance of $\sim$ 10,800 km, representing the upper limit of the corona-disk separation. Eventually, for the high inclination sources, the lag becomes soft, and its magnitude decreases to 10 ms, referring to the inferred separation of $\sim$3,000 km, suggesting a compact coronal region.  
Although these length scales do not directly represent the physical size or the actual distance of the corona, they serve as a tracer of the characteristic scales of the coronal extent. 
Notably, the relative changes in these inferred distances can be used as a proxy to trace the physical evolution of the coronal extent over time. 
Using the recently developed physically motivated time-dependent Comptonisation model, VKOMPTHDK  \citep{2021MNRAS.503.5522K, 2022MNRAS.515.2099B}, \cite{2023MNRAS.520..113R} found a significant contraction of the coronal size from 10,000 km to 3,000 km, as the QPO frequency increases from $\sim$1.8 Hz to 9 Hz in MAXI J1535-571, which is consistent with the implications of our results. 
\cite{2019Natur.565..198K} also suggested that the corona contracts during the hard-to-soft transition, which would account for the soft lags at the higher frequencies for the high inclination sources. Moreover, when the QPO crosses the critical frequency ($\sim$3 Hz) for this source, quenching of the radio emission from the jet was observed \citep{2019ApJ...883..198R}. 
\cite{,2022NatAs...6..577M} observed a similar scenario, where as the QPO frequency increases, the values of the lags decrease and continue to be soft, which was concurrent with a decrease in the observed radio flux. Both results being consistent with a decreasing size of the X-ray corona. 
Based on the result, they proposed that the X-ray corona transforms into the radio jet. 
These interpretations are further supported by the energy-dependent RMS and lag spectra.

At higher frequencies (above $\sim$2 Hz), the RMS increases with energy, resulting in a positive slope. This trend of increasing fractional RMS with energy is well known for BH-LMXBs, suggesting that the Comptonised component is the primary driver behind the variability that generates the QPO, rather than the disk component \citep{2020MNRAS.499..851Z}. 
However, in certain specific scenarios, the RMS has been observed to become saturated or even start to decrease after 10 keV \citep{2023MNRAS.525.4515H,2024MNRAS.527.7136B}. 
Interestingly, in our analysis, for low frequency QPOs observed prior to the hard to HIMS transition, the RMS spectrum exhibits a negative trend (Figure~\ref{Fig:9}), suggesting less variability at high energy. This could be a result of a cold corona located farther away from the central engine, which could dampen the fluctuations.
Notably, the slope of the RMS spectra (right panel of Figures~\ref{Fig:8} and ~\ref{Fig:9}) flips sign near QPO frequency of 2 Hz and hue value of $150^\circ$ marking the hard to HIMS transition, a result reported for the first time using \emph{AstroSat} data. 
During this transition, radio jets emerge \citep{2022NatAs...6..577M}, suggesting a physical transformation of the corona from a compact and active configuration in HIMS to a more extended and possibly cooler structure, potentially driven outward by the launching jet. This far-distant cold corona can also explain the large hard lags observed at lower frequencies.
For the low inclination sources, though the time lag spectra always showed a positive trend (Figure~\ref{Fig:10}), the RMS energy spectra showed a transition from a decreasing to an increasing trend (Figure~\ref{Fig:8}), suggesting that the effects of the transition from hard to HIMS state on the QPO RMS are similar for both high and low inclination sources.
The transition frequency $\sim$2 Hz likely corresponds to the characteristic radius of Lense-Thirring precession (assuming this to be the QPO origin mechanism), beyond which the inner region of the accretion disk reaches such conditions where a transition in the emission behaviour occurs.
The transition of the QPO behaviour at specific hue values represents a spectral state transition. This indicates that the QPO phenomenon is strongly connected to the evolution of the broadband variability contributions from the disk and the corona across the state transition.

However, for explaining the QPO characteristics, certain alternate physical scenarios can also become important. For instance, \citet{Salvesen_2022} 
showed that soft lags may not be solely due to light travel times, but can also comprise a significant contribution from thermalization (scattering) timescales in the disk atmosphere during intermediate and soft states. However, this scenario cannot account for the large soft lags observed in the harder states and, therefore, does not provide a comprehensive explanation of our results.
Additionally, recent X-ray polarization observations \citep{2024ApJ...968...76I} show that there is little to no significant geometrical evolution of the corona during state transitions, which favors an extended corona in the disk plane during the hard-intermediate state. This indicates that the evolution of time lags might involve a combination of geometrical and radiative effects, and that a purely geometrical explanation (based on a specific configuration) of the lag transition might not be fully appropriate. However, constraining the contributions of these distinct effects requires higher-fidelity modelling and rigorous spectro-temporal studies using robust, physical emission models.

In this study, we observed that with increasing QPO frequency, the averaged time lag and the slope of the lag spectra for the low inclination sources remain positive in contrast with the high inclination sources (Figure~\ref{Fig:7}). 
The inclination dependence of the lag behaviour can be understood within the framework of relativistic precession models. At higher frequencies (smaller radii), however, relativistic and geometric effects such as Doppler boosting, projected area variations, and inclination-dependent optical depth—become significant, leading to distinct flux dependency for high and low-inclination systems \citep{2013ApJ...778..165V}. At lower QPO frequencies (larger radii), relativistic effects are relatively weak, where both inclination classes exhibit similar lag behaviour.
\citet{2015MNRAS.447.2059M} suggested 
that the angular dependence of coronal emission depends on the distance from the black hole. Emission from the outer hot-flow regions is more visible in low inclination sources, whereas the inner regions
contribute more strongly in high-inclination systems. This can lead to different lag behaviours for different inclination sources.
Furthermore, it has been proposed that in high-inclination systems, the optical depth along the line of sight varies with the QPO phase, whereas low inclination systems exhibit a consistently lower optical depth across all QPO phases, showing no such significant variation  \citep{2020MNRAS.491..313A}. This can lead to corresponding changes in the observed flux originating from the various components (disk/corona/ reflection), resulting in the distinct phase-lag behaviours observed between high- and low inclination sources.
Overall, while the detailed lag behaviour differs between inclination classes, the consistent trends decreasing lag amplitude, evolution of RMS spectra, and transitions around $\sim$2 Hz point to a common physical origin driven by the evolution of the corona and its coupling with the accretion flow and jet during state transitions.

\subsection{\bf Evolution of the harmonics behaviour}
Most of the harmonic components have been detected from the high inclination sources (except one in low inclination) in our sample during hard and HIMS states, displaying distinct evolutionary behaviour that offers critical insights into the accretion dynamics. The decline in the $Q$ factor of the harmonic with increasing hue (Figure~\ref{Fig:6}) suggests a loss of coherence as the accretion flow becomes more thermally dominated as the system transitions into softer spectral states. 
The increase in harmonic frequency with hue in the HIMS mirrors the primary QPO trend, indicative of a shared or related physical origin \citep{10.1111}.
However, the time lag behaviour of the harmonic is completely different from that of the QPO.  
The harmonic notably shows hard lags only (Figure~\ref{Fig:12}) that remain fairly constant, whereas the QPO lag systematically evolves, changing from hard to soft.
Similar results were previously reported by \cite{2013ApJ...778..136P,2020MNRAS.494.1375Z} for low-frequency QPOs in GRS 1915+105. In our work, by analysing QPOs from multiple high inclination sources, we find that the variation of the lag behaviour of the QPO and its harmonic remains consistent across different systems. 
To explain the contrasting lag behaviour, \citet{2020MNRAS.494.1375Z} proposed that the second harmonic may be generated through a different mechanism or region than the QPO and its sub-harmonic. 
This is further supported by \cite{2016MNRAS.458.1778A}, who found that the energy spectrum of the harmonic in GX 339-4 shows a softer energy spectrum than its QPO. They suggested that QPO originates from an inhomogeneous corona, whereas the harmonic originates from the outer part of the flow; thus, the energy spectrum is softer.
A striking aspect of our QPO-harmonic behaviour is their opposite time lag evolution after the state transition (Figure~\ref {Fig:12}), which was also observed by \citet{2020MNRAS.498.2757G,2023MNRAS.525.4515H}. 
For explaining this, they used the model proposed by \cite{2020MNRAS.498.2757G} and inferred that the variability responsible for the QPO gets generated in the hot inner flow, whereas, for the harmonic, the variability is produced in the thermal disk.
As discussed earlier, if the corona in the hard state is extended, this can result in higher time lag values at the highest energy band for the harmonic as well ($\sim$30 ms, first lower panel of Figure~\ref{Fig:12}). 
The lag spectra of the harmonics remain largely unaffected by the state transition, in contrast to the fundamental component, which exhibits a softening consistent with reverberation effects. This indicates that alterations in the inner coronal region responsible for reflecting photons do not influence the harmonic behaviour. These findings suggest that the harmonics likely originate in the relatively outer and less-evolving regions of the accretion disk.

\section{Summary}
\begin{enumerate}
    \item 
    We employed the power colour analysis using \emph{AstroSat} data of  14 BH-XRBs and demonstrated it to be an effective tool for state identification. This method offers a unified framework for comparing variability across sources, which is challenging using traditional HID diagrams.

    \item 
    We searched for and detected multiple QPOs in our sample and conducted a detailed analysis of the evolution of their properties with hue for the high and low inclination sources.

    \item The time lag of the QPO exhibited a change in its evolution following the QPO frequency of $\sim 2$ Hz for the high inclination sources. 
  
    \item 
    The transition of time lag of the QPO of the high inclination sources from hard to soft during the hard to HIMS transition, concurrent with the change in RMS spectral trend, suggests a significant change in the coronal geometry. Also, the confined soft lag values $\lesssim$10 ms suggest a compact, small corona behind the soft reverberation, whereas before the transition, the large lag value of $\sim$36 ms suggests an extended corona, considering a lamppost-like coronal configuration. However, alternative scenarios involving scattering processes and different coronal geometries may also contribute significantly to the observed soft lags.

    \item 
    We have observed that, though the time lag does not show transition for the low inclination sources, their RMS energy spectral slope changed its sign near the same hue ($\sim$150$^\circ$) and QPO frequency (2 Hz) values where the transition in the time lag behaviour was observed for the high inclination sources.
   
    \item 
    In our work, we extend the analysis beyond individual sources and demonstrate that the transition in time lag behaviour for the QPO and the absence of the same for its harmonic remain consistent across multiple high inclination black hole systems.
    Unlike the fundamental QPO, harmonic lag spectra remain largely unaffected by the state transition, suggesting they might originate in the outer edge of the inner flow and are unaffected by inner coronal changes. 

    \item The change in QPO behaviour at specific hue values, marking spectral state transitions, indicates a strong link between the QPOs and the evolving broadband variability originating from the disk and corona.
    
\end{enumerate}

\section*{Acknowledgements}
This study incorporates data collected through the \emph{AstroSat} mission, which is accessible to the public via the ISRO Science Data Archive for the \emph{AstroSat} Mission. The Indian Space Science Data Centre (ISSDC), ISRO, facilitates user access to the data (https://astrobrowse.issdc.gov.in/
astro$\_$archive/archive/Home.jsp). We are grateful to the LAXPC team for providing the data and the required software tools for the analysis.
We also thank Prof. Ranjeev Misra for his help during the data analysis process.
We extend our appreciation for the support provided by ISRO under the \emph{AstroSat} Archival Data Utilization Programme (DS 2B-13013(2)/4/2020\-Sec.2). 

\section*{DATA AVAILABILITY}
Data analysed in this work is publicly available on the Indian Space Science Data Center (ISSDC) website(https://astrobrowse.
issdc.gov.in/astroarchive/archive/Home.jsp).

\appendix

\bibliographystyle{mnras}
\bibliography{example} 



\end{document}